\documentclass[journal]{IEEEtran}
\pdfoutput=1
\usepackage{amsfonts}
\usepackage{cite}
\usepackage{balance}
\usepackage{graphicx,color,overpic,psfrag}
\usepackage{amsmath}
\usepackage{times}
\usepackage{latexsym}
\usepackage{bm}
\usepackage{amssymb}
\usepackage{cases}
\usepackage{array}
\usepackage{fancyhdr}
\usepackage{stfloats}
\usepackage{setspace}
\usepackage{booktabs}
\usepackage{float}
\usepackage{graphicx}
\usepackage{lettrine}

\usepackage{epstopdf}
\usepackage{pifont}
\usepackage{xcolor}
\usepackage{colortbl,booktabs}
\usepackage{tabu}
\usepackage{subfig}
\usepackage{multirow}

\begin{document}
%
\title{Implementation of Massive MIMO Uplink Receiver on RaPro Prototyping Platform}

\author{Xuanxuan~Gao,~\IEEEmembership{Student Member,~IEEE,}
        Zhichao~Huang,~\IEEEmembership{Student Member,~IEEE,}
        Xintong~Lu,~\IEEEmembership{Student Member,~IEEE,}
        Senjie~Zhang,~\IEEEmembership{Member,~IEEE,}
        Chao-Kai~Wen,~\IEEEmembership{Member,~IEEE,}
        and 
        Shi~Jin,~\IEEEmembership{Senior~Member,~IEEE}}


\maketitle

\begin{abstract}
The updated physical layer standard of the fifth generation wireless communication suggests the necessity of a rapid prototyping platform. To this end, we develop RaPro, a multi-core general purpose processor-based massive multiple-input-multiple-output (MIMO) prototyping platform. To enhance RaPro, high performance detection and beamforming are needed, whereas both of them request for accurate channel state information (CSI). In this paper, linear minimum mean square error (LMMSE)-based channel estimator is adopted and encapsulated inside RaPro to gain more accurate CSI. Considering the high comlexity and unknown of channel statistics, we design low-complexity LMMSE channel estimator to alleviate the rising complexity along with increasing antenna number and set more computational resource aside for massive MIMO uplink detection and downlink beamforming. Simulation results indicate the high mean square error performance and robustness of designed low-complexity method. Indoor and corridor scenario tests show prominent improvement in bit error rate performance. Time cost analysis proves the practical use and real-time transmission ability of the implemented uplink receiver on RaPro. 
\end{abstract}

\section{Introduction}

\lettrine[lines=2]{M}{assive} multiple-input multiple-output (MIMO) is a potentially disruptive technology in the fifth generation (5G) cellular network [1, 2]; it utilizes a large excess of base station (BS) antennas compared with user equipment (UEs). The combination of massive MIMO and the promising transmission technology and orthogonal frequency division multiplexing (OFDM) enables full usage of the spatial degrees of freedom and leads to high data transmission rate. The massive MIMO system with an accurate channel estimator cannot only provide channel state information (CSI) to the downlink of a time-division duplex (TDD) massive MIMO system to support multi-stream transmission through beamforming but is also necessary in state-of-the-art MIMO detectors. 
 

The aforementioned system encounters limitations in uplink receiver implementation, especially for computational complex operations, channel estimation, and MIMO detection. The complexity of channel estimation module is connected with number of antennas and subcarriers. Consider, for example, a 128$\times$12 massive MIMO system with 1200 subcarriers used to transmit data. For such a system, there are 128$\times$12 times $(1200\times1200)\times(1200\times1)$ complex matrix-vector multiplication to be computed within one slot when adopting linear minimum mean square error (LMMSE) channel estimator. For the general purpose processor, such amount of computation cannot be down within one slot, which means the detection process has to be delayed and the transmission is not real-time. To enable real-time transmission in massive MIMO system, employing low complexity channel estimators and saving more computational resource for detection and beamforming plays a critical role. Such condition suggests a reduction in LMMSE estimation using the nearest taps in [3]; similarly, a low-rank LMMSE channel estimator was proposed in [4], and an analyzed fixed design exhibited robustness to changes in channel correlation and signal-to-noise ratio (SNR). A novel low complexity LMMSE channel estimator partitions the channel autocorrelation matrix into small sub-matrices in non-overlapping and overlapping manners, significantly reducing complexity [5]. For the MIMO detector, MMSE detector presents the best performance in linear detection and is first considered during implementation. Implementation and verification of the receiver with linear channel estimation schemes and MMSE MIMO detection technique are field-programmable gate array (FPGA)-based and are extremely time-consuming considering the development cycle. The Application Specific Integrated Circuit design of the low-complexity LMMSE channel estimator in [6] is constrained by a specialized hardware, instead of being implemented and verified by over-the-air (OTA) measurements in a general communication system testbed.

In [7], we have developed RaPro, which is a novel prototyping platform that combines FPGA-privileged modules from a software-defined radio (SDR) platform and high-level programming language for advanced algorithms from a server with multi-core general purpose processors (GPPs). Based on this platform, data processing algorithms of a massive MIMO uplink receiver, which includes channel estimation and MIMO detection, can be implemented and verified rapidly. The present paper focuses on implementation of receiver data processing, which includes low-complexity channel estimation and MIMO detection schemes for the uplink long term evaluation (LTE)-like TDD massive MIMO systems on the RaPro server. Monte-Carlo simulation and OTA measurements are conducted to evaluate the performance of the uplink receiver.

The subsequent contents are organized as follows. Section \uppercase\expandafter{\romannumeral2} establishes the system model to provide an overview of the massive MIMO system, which is in accordance with the prototype system RaPro. Section \uppercase\expandafter{\romannumeral3} discusses the top design of uplink receiver on multi-core GPPs of RaPro after set up. Section \uppercase\expandafter{\romannumeral4} presents the specific implementation procedures of practical channel estimation and MIMO detection schemes after elaborating the designed mapping rules 3W-LMMSE and 12W-LMMSE. Section \uppercase\expandafter{\romannumeral5} presents and discusses simulation and OTA measurement results. Section \uppercase\expandafter{\romannumeral6} provides the concluding remarks.

\emph{Notation:} Matrices and vectors are denoted by uppercase and lowercase letters in boldface, respectively. An identity matrix is denoted by ${\bf{I}}$ or ${\bf{I}_N}$ when specifying its dimension $N$ is necessary; $(\cdot)^{\rm{H}}$, $(\cdot)^{\rm{T}}$ and $(\cdot)^{-1}$ stand for the conjugate transpose, transpose and inverse operations respectively; $E\left\{  \cdot  \right\}$ is the statistical expectation; $\left\lfloor  a  \right\rfloor $ denotes the highest integer no larger than $a$; $*$ denotes convolution operation; $A \buildrel \Delta \over = B$ means $B$ as $A$; and $diag( \cdot )$ represents a diagonal matrix whose diagnal elements are the vector inside the brackets.

\section{System Model}
\begin{figure}[!t]
	\centering
	\includegraphics[width=3.3in]{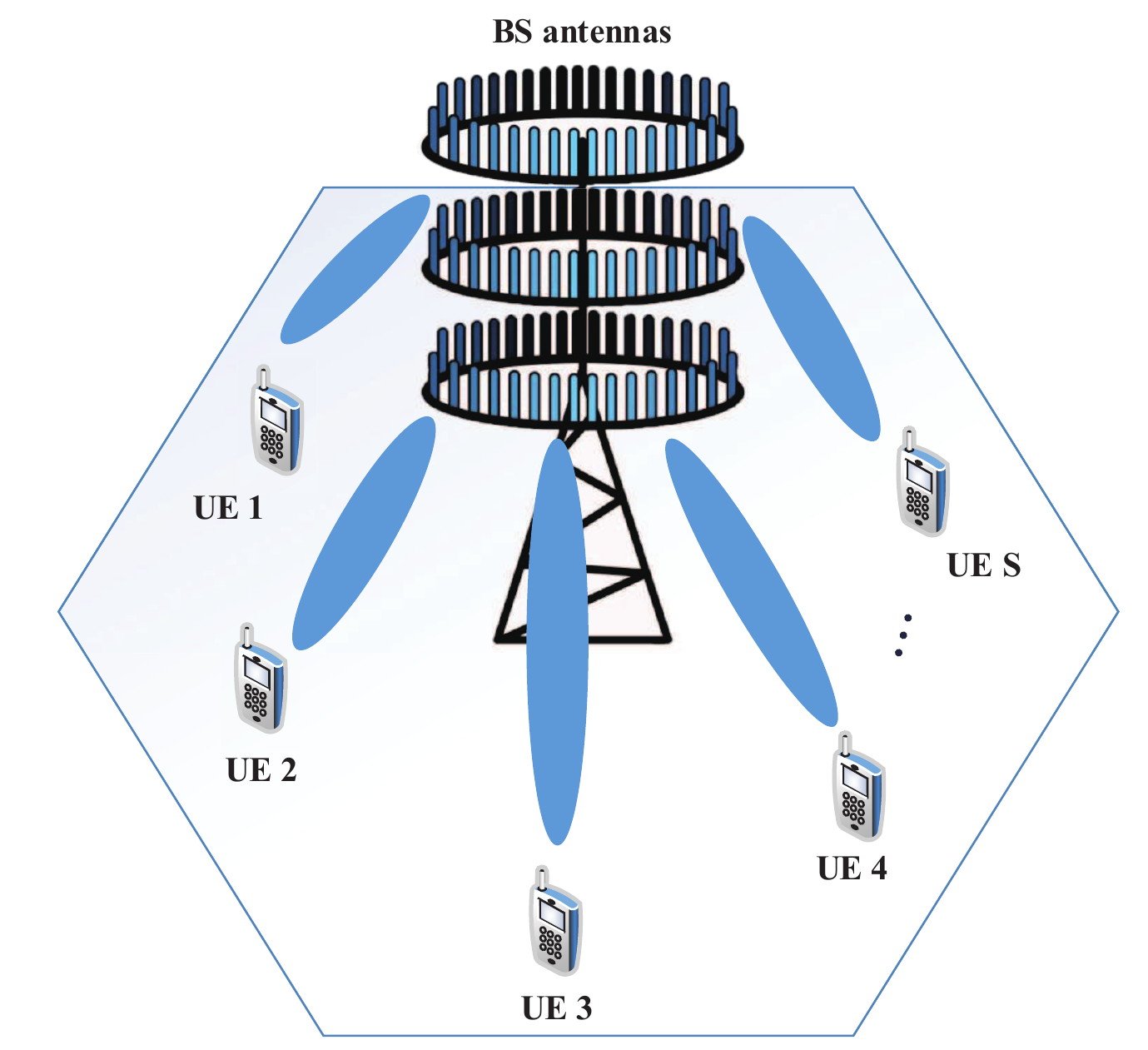}
	\caption{ Illustration of a single-cell MU massive MIMO system. In this system, the BS equipped with $R$ antennas serves as $S$ single-antenna UEs. Data transmission from UE to BS occurs in an uplink chain and a reversed downlink.}
	\label{figMIMO}
\end{figure}

\begin{figure}[!t]
	\centering
	\includegraphics[width=3.5in]{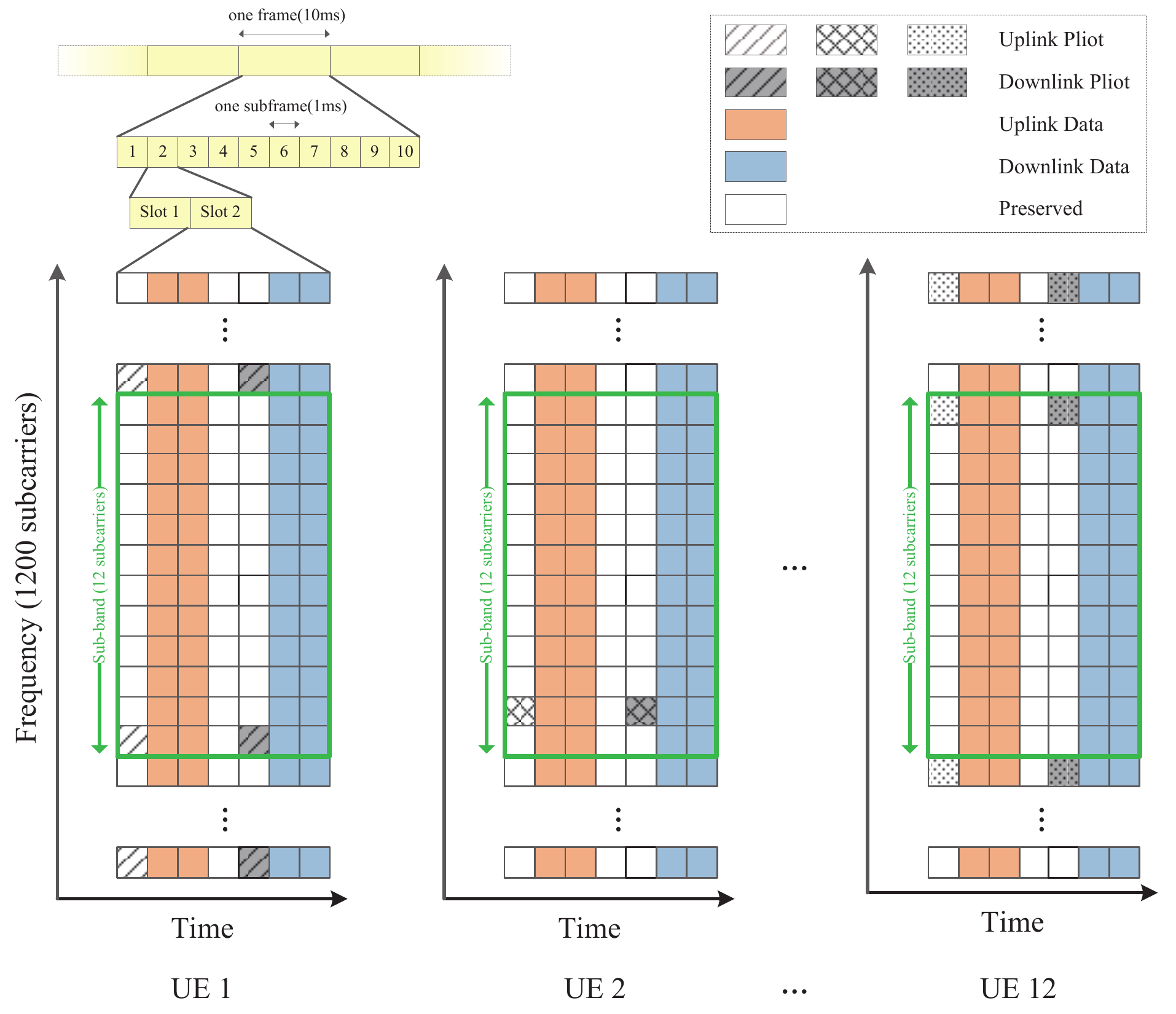}
	\caption{Time-frequency resource grids for 12 single-antenna UEs. Allocation of pilots inside a sub-band corresponds to individual UEs in Subframes 2-10, whereas Subframe 1 is filled with primary synchronization signals for synchronization.}
	\label{fig1}
\end{figure}

Figure \ref{figMIMO} illustrates the uplink chain of a single-cell TDD-based multi-user (MU) massive MIMO system with OFDM transmission scheme. As shown in the figure, BS is equipped with ${R}$ antennas and serves ${S}$ single-antenna UEs. For one antenna to another antenna transmission link, the channel is assumed to be a $M$ path multipath channel, and time delay of each path is ${\tau _m}$, where $1 \le m \le M$. At moment ${t_0}$, the receiving signal on a specific subcarrier is determined as follows:  
\begin{equation}\label{eq1}
y\left( {{t_0}} \right) = \sum\limits_{m = 0}^{M - 1} {h\left( {{\tau _m}} \right)} x\left( {{t_0} - {\tau _m}} \right) + n\left( {{t_0}} \right).
\end{equation}

In \eqref{eq1}, ${x\left( {{t_0} - {\tau _m}} \right)}$ corresponds to the transmitting signal, $n\left( {{t_0}} \right)$ is additive white Gaussian noise (AWGN), and ${h\left( {{\tau _m}} \right)}$ denotes the channel impulse response. Formula \eqref{eq1} is also the form of convolution. Thus, we obtain the following:
\begin{equation}\label{eq2}                                    
	y\left( {{t_0}} \right) = h\left( {{t_0}} \right) * x\left( {{t_0}} \right) + n\left( {{t_0}} \right).
\end{equation}

For the ${r^{th}}$ receiving antenna, where $1 \le r \le R$, the receiving signal corresponds to the superposition of all UE signals through channels:
\begin{equation}\label{eq3}                                    {y_r}\left( {{t_0}} \right) = \sum\limits_{s = 1}^S {{h_{r,s}}\left( {{t_0}} \right) * {x_s}\left( {{t_0}} \right) + {n_r}\left( {{t_0}} \right)}.
\end{equation}

When pilot transmitting signals are considered, Formula \eqref{eq3} can only feature the influence of one UE by setting the pilot properly. Suppose only the ${s^{th}}$ $\left( {1 \le s \le S} \right)$ UE sends the pilot to a certain moment and subcarrier, whereas the others send zero-value signals, Formula \eqref{eq3} then transforms into the following:
\begin{equation}\label{eq4}                                                
{y_r}\left( {{t_0}} \right) = {h_{r,s}}\left( {{t_0}} \right) * {x_s}\left( {{t_0}} \right) + {n_r}\left( {{t_0}} \right).
\end{equation}

The signal relationship in the above formula can be observed in the frequency domain, with most channel estimators being subsequently conducted. Formula \eqref{eq4} is transformed from a time domain into a frequency domain using fast Fourier transform (FFT) and becomes the following:
\begin{equation}\label{eq5}                                                
{Y_r}\left( k \right) = {H_{r,s}}\left( k \right){X_s}\left( k \right) + {N_r}\left( k \right).
\end{equation}

In Formula \eqref{eq5}, ${Y_r}\left( k \right)$, ${X_s}\left( k \right)$, ${N_r}\left( k \right)$ and ${H_{r,s}}\left( k \right)$ denote the receiving signal of ${r^{th}}$ antenna, transmitting signal of ${s^{th}}$ UE, AWGN noise of ${r^{th}}$ antenna, and the current channel in ${k^{th}}$ subcarrier in the frequency domain, respectively. A total of ${N_{FFT}}$ subcarriers and ${{N_{FFT}}-N}$ zero-padding subcarriers are observed; $N$ denotes the number of subcarriers in the recource block of an OFDM signal. Figure \ref{fig1} illustrates the resource allocation of 12 UE systems similar to those in [8]. The pilot of each UE gradually occupies a subcarrier in the sub-band, and this process is then repeated. For example, the pilots of UE 1 occupy subcarrier $1$, ${1+S}$, ${1+2S}$, $ \cdots $, and ${1+\left\lfloor {\frac{N}{S}} \right\rfloor }$. Therefore, for ${s^{th}}$ UE, additional formulas similar to Formula \eqref{eq5} can be derived in vector and matrix form:
\begin{equation}\label{eq6}                                                
{{\bf{Y}}_r}{\bf{ = }}{{\bf{X}}_s}{{\bf{H}}_{r,s}}{\bf{ + }}{{\bf{N}}_r},
\end{equation}
where\\
${{\bf{X}}_s} \buildrel \Delta \over = {\left[ {\begin{array}{*{20}{c}}
		{diag\left( {{X_s}\left( {{k_1}} \right),{X_s}\left( {{k_2}} \right) \cdots {X_s}\left( {{k_K}} \right)} \right)}
		\end{array}} \right]_{K \times K}},$\\
${{\bf{Y}}_r} \buildrel \Delta \over = {\left[ {\begin{array}{*{20}{c}}
		{{Y_r}\left( {{k_1}} \right)}&{{Y_r}\left( {{k_2}} \right)}& \cdots &{{Y_r}\left( {{k_K}} \right)}
		\end{array}} \right]^{\rm{T}}},$\\
${{\bf{H}}_{r,s}} \buildrel \Delta \over = {\left[ {\begin{array}{*{20}{c}}
		{{H_{r,s}}\left( {{k_1}} \right)}&{{H_{r,s}}\left( {{k_2}} \right)}& \cdots &{{H_{r,s}}\left( {{k_K}} \right)}
		\end{array}} \right]^{\rm{T}}},$\\
${{\bf{N}}_r} \buildrel \Delta \over = {\left[ {\begin{array}{*{20}{c}}
		{{N_r}\left( {{k_1}} \right)}&{{N_r}\left( {{k_2}} \right)}& \cdots &{{N_r}\left( {{k_K}} \right)}
		\end{array}} \right]^{\rm{T}}},$\\
and ${k_1}$, ${k_2}$, $\cdots$, and ${k_K}$ represent $K$ pilot tones $s$, $s + S$, $\cdots$, and ${s + \left\lfloor {\frac{N}{S}} \right\rfloor }$, respectively. The pilot signal is assumed to satisfy the condition that $E\left\{ {{X_s}\left( k \right)} \right\} = 0$ and $Var\left\{ {{X_s}\left( k \right)} \right\} = {\sigma_s ^2}$ and the white Gaussian independent and identically distributed (i.i.d) noise component features a mean of $E\left\{ {{N_r}\left( k \right)} \right\} = 0$ and a variance of $Var\left\{ {{N_r}\left( k \right)} \right\} = \sigma ^2,k = {k_1},{k_2}, \cdots ,{k_K}$. 

Formula \eqref{eq6} describes the system in terms of OFDM architecture. When spatial multiplexing MIMO properties are considered, the system can be formulated as follows. For each OFDM subcarrier, $S$ transmitting symbols and $R$ receiving symbols are detected. Thus, the system is an $R \times S$ MIMO system. Considering subcarrier $k$, the system function is as follows:
\begin{equation}\label{eq7}
{\bf{Y}} = {\bf{HX}} + {\bf{N}},
\end{equation}
where\\
${\bf{X}} \buildrel \Delta \over = {\left[ {\begin{array}{*{20}{c}}
		{{X_1}\left( {{k_k}} \right)}&{{X_2}\left( {{k_k}} \right)}& \ldots &{{X_S}\left( {{k_k}} \right)}
		\end{array}} \right]^{\rm{T}}},$\\
${\bf{Y}} \buildrel \Delta \over = {\left[ {\begin{array}{*{20}{c}}
		{{Y_1}\left( {{k_k}} \right)}&{{Y_2}\left( {{k_k}} \right)}& \ldots &{{Y_R}\left( {{k_k}} \right)}
		\end{array}} \right]^{\rm{T}}},$\\
${\bf{N}} \buildrel \Delta \over = {\left[ {\begin{array}{*{20}{c}}
		{{N_1}\left( {{k_k}} \right)}&{{N_2}\left( {{k_k}} \right)}& \ldots &{{N_R}\left( {{k_k}} \right)}
		\end{array}} \right]^{\rm{T}}},$\\
and ${\bf{H}}$ denote the $R \times S$ channel matirx, in which the $(r,s)$ element corresponds to ${H_{r,s}}\left( {{k_k}} \right)$. 
\begin{figure}[H]
	\centering
	\includegraphics[width=.5\textwidth]{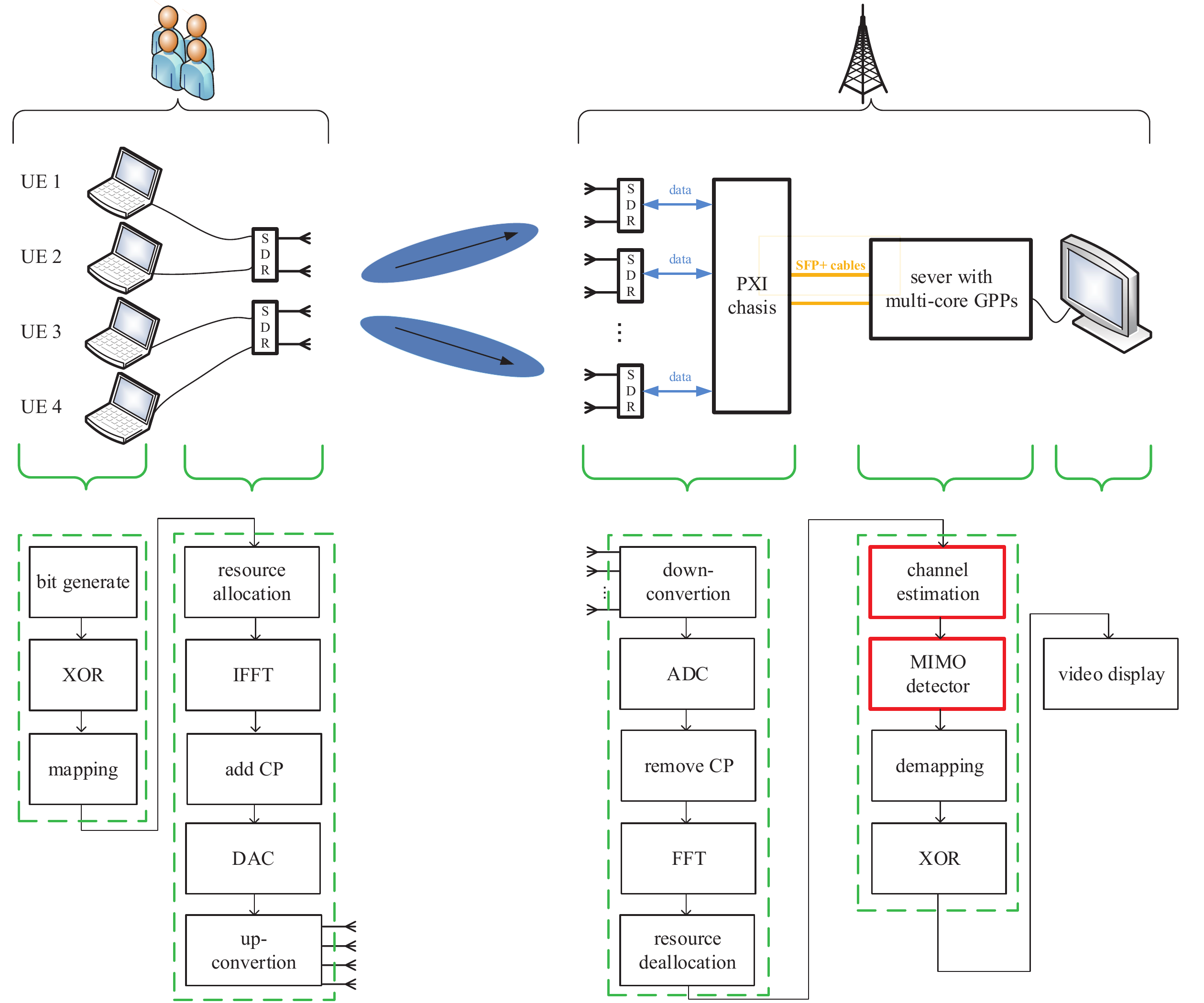}
	\caption{Architecture of RaPro and the function of each module. RaPro consists of SDRs contained in the FPGA module and a server with multi-core GPPs. FPGA-privileged operations, such as FFT/IFFT, are conducted in SDRs. Computational complex algorithms similar to channel estimation and MIMO detection are implemented in the server.}
	\label{fig2}
\end{figure}


\section{Overall Design of Uplink Receiver on RaPro}
RaPro [7] is a novel 5G rapid prototyping testbed recently proposed by our team; it utilizes FPGA in SDR devices and GPPs in a server. This section describes a RaPro setup to present an overview of this prototyping platform. This portion also shows the overall design of an uplink receiver to describe the multi-core GPP-based multi-thread paralleled data processing procedure and describes function calling and a typical data format.

\subsection{RaPro Setup}
Figure \ref{fig2} shows the architecture of RaPro and the corresponding function of each module. RaPro can be configured as a duplex mode. In this paper, we consider the uplink chain to elaborate on the implementation of uplink receiver in BS.

\begin{figure*}[!t]
	\centering
	\includegraphics[width=.97\textwidth]{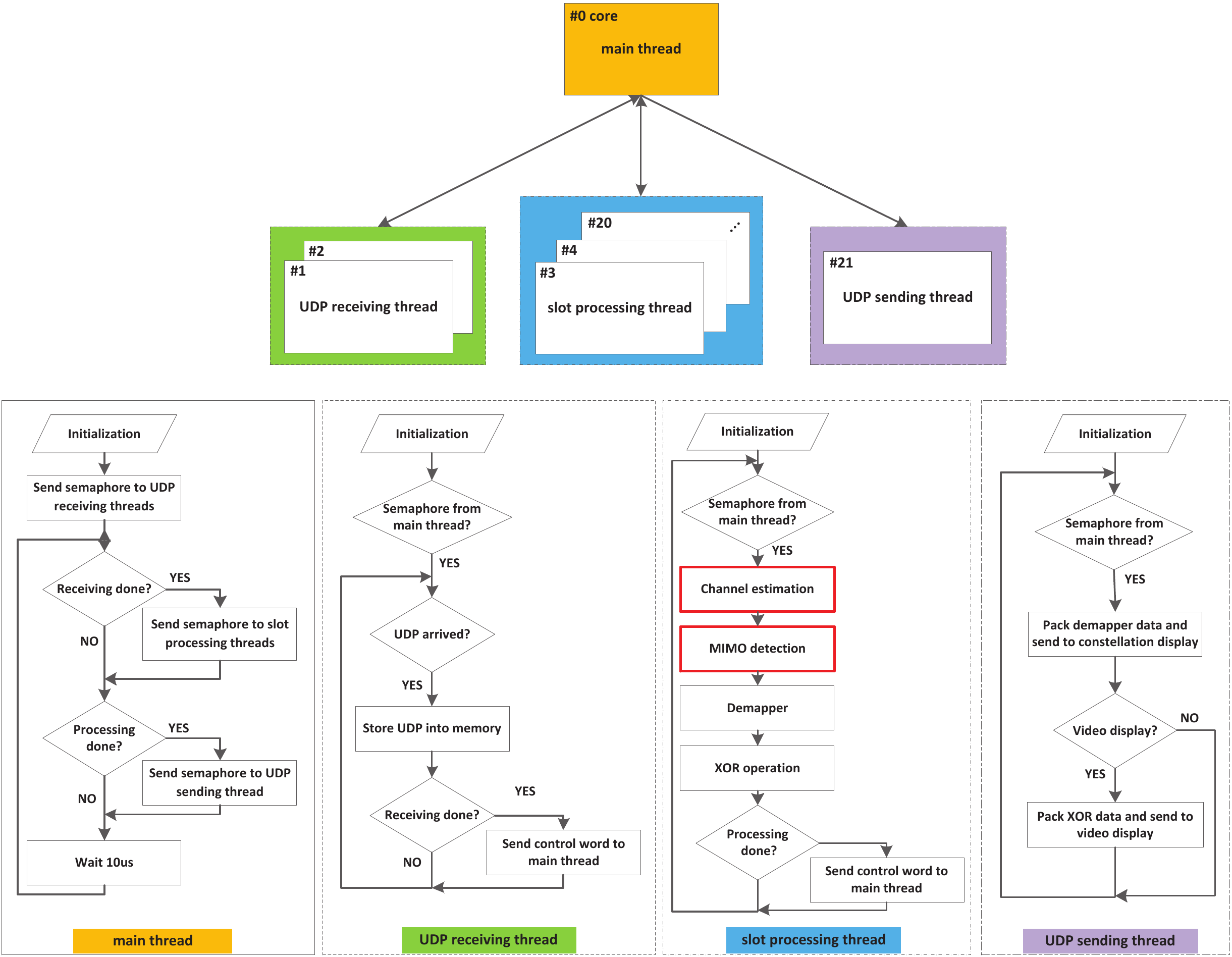}
	\caption{Thread-interactive relationship diagram and flowchart of each thread. The bottom three threads are controlled by the main thread. Receiving and processing threads send control words back to the main thread to mark their finish. From the main thread, the receiver starts and activates two UDP-receiving threads. A corresponding slot processing thread runs once a slot of UDP data has been received. Otherwise, the thread waits for 10 us and enters the next loop directly. The UDP-receiving thread waits for UDP, stores UDP into memory, sends control word to the main thread, and keeps receiving more UDP. Baseband data processing, which includes channel estimation, MIMO detection, and demapping, are conducted in slot processing threads. After processing, the processing thread sends a control word to the main thread and enters the next loop. The recovered data in the UDP-sending thread are packed and sent to a laptop to show the constellation. When data comprise video stream, the UDP-sending thread will send XOR data to video display.}
	\label{fig3}
\end{figure*}

On the transmitter side of RaPro uplink transmission, four active UE strings of data bits are generated and randomized by four hosts and mapped to a constellation based on quadrature amplitude modulation (QAM) schemes. Resource allocation, inverse FFT (IFFT)-based OFDM modulation, and cyclic prefix (CP) adding are conducted in two SDR nodes NI USRP-2943R. After analog-to-digital conversion and up-conversion, UE data are transmitted in a simplified LTE-like 10 millisecond (ms) radio frame structure (Figure \ref{fig1}) and are sent through four individual dipole antennas with 40 MHz bandwidth and 4.1 GHz\footnote{The center frequency configuration is adjustable in the range of 1.2GHz-6GHz when using corresponding antennas.} center frequency. 

On the receiver side of RaPro uplink transmission, wireless signals are collected through an 8$\times$2 uniform planar array connected with eight SDRs to conduct an FPGA-privileged operation, which includes CP removal, FFT-based OFDM demodulation, and resource deallocation. The obtained 16 baseband symbol streams are aggregated subsequently through a PCI eXtensions for Instrumentation (PXI) chassis that contains NI 6592 in the form of User Datagram Protocol (UDP) packages. These UDP packages are sent through small form-factor pluggable plus (SFP+) cables from 10 G Ethernet ports to the multi-core server that contains 20 Intel Xeon E5-2680 v2 @ 2.8 GHz processors. The baseband signal processing algorithms of the receiver include channel estimation, MIMO detection, demapping, and XOR and are implemented on the server; the algorithms are further extended in the next subsection. Raw bits are sent to a laptop to display the constellation or play a video. Accurate timing synchronization is ensured among UEs or BS antennas through PXIe-6674T inside the PXI chassis, whereas air interface synchronization is achieved by a fine-timing synchronization algorithm implemented in SDRs.

\begin{figure*}[!t]
	\centering
	\includegraphics[width=.97\textwidth]{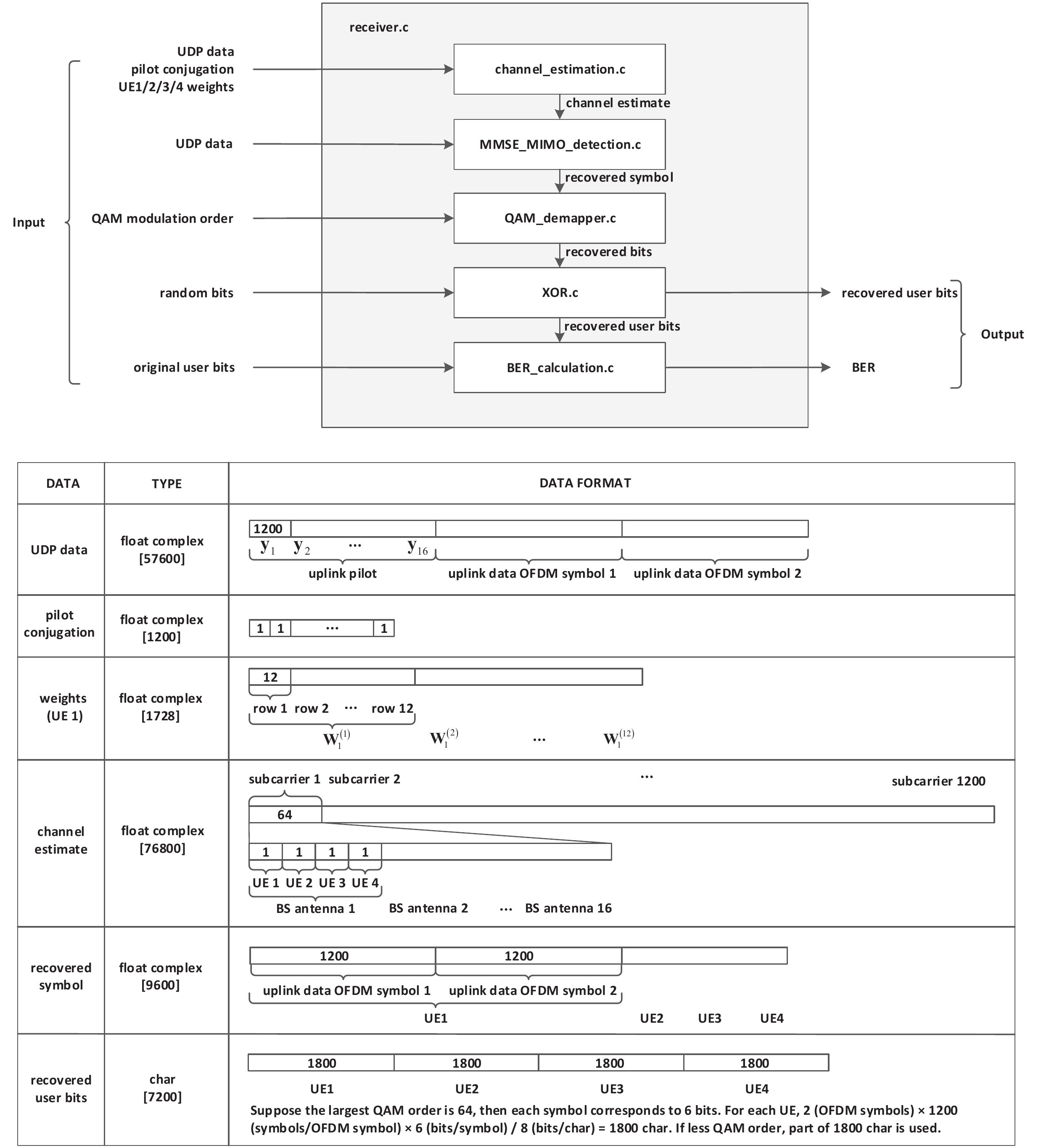}
	\caption{Function calling and data format. An active slot processing thread calls receiver.c to conduct data processing. Five subfunctions channel\_estimation.c, MMSE\_MIMO\_detection.c, QAM\_demapper.c, XOR.c, and BER\_calculation.c in receiver.c should be called in sequence to realize the corresponding functions as the literal meanings of function names. Inputs and outputs of each function should follow the prescriptive layout to enable a functional design. A typical example of data format is given, and the numbers inside the rectangles refer to the number of data in the form of corresponding types. All the data presented are stored in a one-dimensional dynamic array, allowing  Intel MKL functions to conduct matrix multiplication and other operations.}
	\label{fig4}
\end{figure*}

RaPro replaces proprietary hardware with a server that contains GPPs, and this multi-core GPP-based architecture enables real-time, flexible, and scalable implementation of baseband processing algorithms. The server is set up with the Ubuntu operating system and Eclipse platform, and C language is used to perform paralleled threading programming with the aid of Intel Math Kernel Library (Intel MKL) [9], which provides a wide range of optimized functions to calculate and solve mathematical problems, such as complex multiplication between matrices and finding solutions to linear equations. 

\subsection{Top Design of Uplink Receiver}
In the multi-core GPP-based receiver design, RaPro uses multi-threads to perform paralleled processing and binds each thread to one processor core to diminish the context switch overhead. Figure \ref{fig3} summarizes the overall data processing program diagram of the massive MIMO uplink receiver on the server and shows four kinds of threads bound to 22 cores that run in parallel in total: 1 \textbf{main thread} bound to Core 0 for scheduling the other threads, 2 \textbf{UDP receiving threads} bound to Cores 1-2 for receiving UE data from two SFP+ cables, 18 \textbf{slot processing threads} bound to Cores 3-20 for conducting the baseband signal processing to recover UE data bits, and 1 \textbf{UDP sending thread} bound to Core 21 for sending the recovered data to a laptop to display the constellation and video (when UEs send video streams). A total of 18 slot processing threads are considered because Subframe 1 carries synchronization signals, whereas Subframes 2-10 carry UE data (Figure \ref{fig1}). Thus, a 10 ms radio frame includes 18 slot data that must be processed. The same kind of threads can be generated through a self-defined structure using C language.

Figure \ref{fig3} presents the logical relationship among the above four kinds of threads and their specific flow diagrams. Three types of thread are controlled by the main thread, and UDP receiving thread and slot processing thread feedback control words to the main thread to mark their finish. A spinlock is used to ensure that only one thread can access the same memory region at the same time. The program initiates from the main thread, sends semaphore to two UDP receiving threads using the function ``sem\_post,'' and assesses whether the UDP data have been received. Once the procedure is completed, the semaphore is sent to each slot-processing thread. Otherwise or after sending semaphore, the program checks whether slot processing threads have been completed, sends the semaphore to a UDP-sending thread, and waits for 10 microsecond to check the receiving state again. Otherwise, the program waits for 10 microsecond and enters the next loop directly. In the UDP-receiving thread, the program waits for the semaphore from the main thread using the function ``sem\_wait.'' When the semaphore is positive, the program waits until UDP arrives and then stores UDP into memory. Each UDP-receiving thread collects UDP packages from eight receiving antennas and sends a control word to the main thread when the first three uplink OFDM symbols, which include one uplink pilot OFDM symbol and two uplink data OFDM symbols within a slot of eight antennas, are stored. Afterward or when UDP packages are still not ready, the UDP-receiving thread continually checks the arrival of UDP. The received UDP data are transferred to slot processing threads to conduct baseband data processing, such as channel estimation, MIMO detection, and demapping, which is the most vital operation to run the uplink receiver; this topic will be further expounded later. After processing, slot processing threads send control words to the main thread and wait for the next semaphore. The UDP-sending thread packs the recovered data and sends them to the laptop to display the constellation. XOR data are played using VLC when the data include video streams.

In the slot processing thread, operations, such as channel estimation, MIMO detection, demapper, XOR operation, and bit error rate (BER) calculation, are all packed into functions, and these five functions are compressed into one function ``receiver.c.'' An active thread calls receiver.c to conduct data processing (Figure \ref{fig4}). This function calling design provides remarkable benefits. Active slot processing threads share the same code ``receiver.c'' and use different data to conduct parallel data processing. The novel algorithm of channel estimation or MIMO detection that needs verification can be written into a separate function and replace the original one. For a shared code, the unified format of input and output data plays an important role. Figure \ref{fig4} illustrates a typical example of data layout. All the data presented are stored in a one-dimensional dynamic array using the C language command ``malloc.'' This format properly allows Intel MKL functions to conduct matrix multiplication and other operations. Notably, the dynamic arrays must be freed in the end to release memory.

\section{Receiver Implementation on RaPro}
This section discusses implementation of the most computational complex data processing algorithms of a massive MIMO uplink receiver; these algorithms include channel estimation schemes and MMSE MIMO detection. Apart from conventional algorithms, two simplified LMMSE-based channel estimation methods, 3W-LMMSE and 12W-LMMSE channel estimation, are designed and explained in detail.

\subsection{Basics of Conventional Channel Estimators and Detectors}
The conventional channel estimation schemes considered include LS channel estimation and LMMSE channel estimation. MIMO detection schemes include ZF detection and MMSE detection. The methods in [10] are shown as follows.

\subsubsection{LS Channel Estimators}
The channel estimator estimates channel ${{\bf{H}}_{{{r,s}}}}$ in Formula \eqref{eq6} and yields the channel estimates ${{{\bf{\hat H}}}_{r,s}}$. LS estimator for the channel from ${s^{th}}$ UE to ${r^{th}}$ antenna on BS is as follows:
\begin{equation}\label{eq8}                                                
{{{\bf{\hat H}}}_{LS,r,s}} = {\bf{X}}_s^{ - 1}{{\bf{Y}}_r}.
\end{equation}
For each pilot tone of the ${s^{th}}$ UE, the value of LS channel estimate is as follows:
\begin{equation}\label{eq9}                                                
{{\hat H}_{LS,r,s}} = \frac{{{Y_r}\left( k \right)}}{{{X_s}\left( k \right)}},k = {k_1},{k_2}, \cdots ,{k_K}.
\end{equation}

LS channel estimate can also be interpolated as an interpolated LS channel estimator through linear interpolation, second-order polynomial interpolation, and cubic spline interpolation [10].

\subsubsection{LMMSE Channel Estimators}
The interpolated LS channel estimator and LMMSE-based channel estimators can be dedicated to make the channel estimate ``dense'' and filled with all the subcarriers. The LMMSE channel estimator is a weighted LS estimator:
\begin{equation}\label{eq12} 
{{{\bf{\hat H}}}_{LMMSE,r,s}} = {{\bf{W}}_{r,s}}{{{\bf{\hat H}}}_{LS,r,s}},
\end{equation} 
where ${{\bf{W}}_{r,s}}$ refers to the weight matrix. The LMMSE channel estimation method determines the optimal weight matrix ${{\bf{W}}_{r,s}}$ by minimizing the mean square error. [10] indicates that the optimal weight matrix ${{\bf{W}}_{r,s}}$ can be written as follows:
\begin{equation}\label{eq13}
{{\bf{W}}_{r,s}} = {{\bf{R}}_{{{\bf{H}}_{r,s}}{{{\bf{\hat H}}}_{LS,r,s}}}}{\left( {{{\bf{R}}_{{{\bf{H}}_{r,s}}{{\bf{H}}_{r,s}}}} + \frac{{\sigma ^2}}{{{\sigma _s^2}}}{\bf{I}}} \right)^{ - 1}},
\end{equation}
where
\begin{equation}\label{eq14}
{{\bf{R}}_{{{\bf{H}}_{r,s}}{{{\bf{\hat H}}}_{LS,r,s}}}} = E\left\{ {{{\bf{H}}_{r,s}{\bf{\hat H}}}_{LS,r,s}^{\rm{H}}} \right\},
\end{equation}
\begin{equation}\label{eq15}
{{\bf{R}}_{{{\bf{H}}_{r,s}}{{\bf{H}}_{r,s}}}} = E\left\{ {{{\bf{H}}_{r,s}}{\bf{H}}_{r,s}^{\rm{H}}} \right\}.
\end{equation}
Formula \eqref{eq13} shows that prior information, such as channel frequency correlation and real-time SNR value, are necessary to conduct LMMSE estimation. However, obtaining such information presents difficulty in practice.
 
\begin{table}[!t]
	\renewcommand{\arraystretch}{1.5}
	\caption{3W-LMMSE Group for UE 1}
	\label{tabel1}
	\centering
	\begin{tabu}[width=2*\textwidth]{llll} 
		\tabucline[2pt]{-}
		\textbf{Index} & \textbf{Subcarrier of}  & ${{{\bf{W}}_{1}}}$ & \textbf{Subcarrier of} \\
		& ${{{\bf{\hat H}}}_{LMMSE,r,s}}$ &   & ${{{\bf{\hat H}}}_{LS,r,s}}$ \\
		\hline
		1 & ${\left( {1,5,9} \right)^{\rm{T}}}$ & ${{\bf{W}}^{\left( 1 \right)}}$ & ${\left( {1,13,25} \right)^{\rm{T}}}$\\
		2 & ${\left( {13,17,21} \right)^{\rm{T}}}$ & ${{\bf{W}}^{\left( 2 \right)}}$ & ${\left( {1,13,25} \right)^{\rm{T}}}$\\
		3 & ${\left( {25,29,33} \right)^{\rm{T}}}$ & ${{\bf{W}}^{\left( 2 \right)}}$ & ${\left( {13,25,37} \right)^{\rm{T}}}$\\
		4 & ${\left( {37,41,45} \right)^{\rm{T}}}$ & ${{\bf{W}}^{\left( 2 \right)}}$ & ${\left( {25,37,49} \right)^{\rm{T}}}$\\
		$ \vdots $ & $ \vdots $ & $\vdots$ & $\vdots$\\
		99 & ${\left( {1177,1181,1185} \right)^{\rm{T}}}$ & ${{\bf{W}}^{\left( 2 \right)}}$ & ${\left( {1165,1177,1189} \right)^{\rm{T}}}$\\
		100 & ${\left( {1189,1193,1197} \right)^{\rm{T}}}$ & ${{\bf{W}}^{\left( 3 \right)}}$ & ${\left( {1165,1177,1189} \right)^{\rm{T}}}$\\
		\tabucline[2pt]{-}
	\end{tabu}
\end{table}

\begin{table}[!t]
	\renewcommand{\arraystretch}{1.5}
	\caption{12W-LMMSE Group for UE 1}
	\label{tabel2}
	\centering
	\begin{tabu}{llll}	
		\tabucline[2pt]{-}
		\textbf{Index} & \textbf{Subcarrier of}  & ${{{\bf{W}}_{1}}}$ & \textbf{Subcarrier of} \\
		&  ${{{\bf{\hat H}}}_{LMMSE,r,s}}$ &  &  ${{{\bf{\hat H}}}_{LS,r,s}}$\\
		\hline
		1 & ${\left( {1:1:12} \right)^{\rm{T}}}$ & ${\bf{W}}_1^{\left( 1 \right)}$ & ${\left( {1:12:133 } \right)^{\rm{T}}}$\\
		2 & ${\left( 13:1:24 \right)^{\rm{T}}}$ & ${\bf{W}}_1^{\left( 2 \right)}$ & ${\left( {1:12:133 } \right)^{\rm{T}}}$\\
		3 & ${\left( 25:1:36 \right)^{\rm{T}}}$ & ${\bf{W}}_1^{\left( 3 \right)}$ & ${\left( {1:12:133 } \right)^{\rm{T}}}$\\
		4 & ${\left( 37:1:48 \right)^{\rm{T}}}$ & ${\bf{W}}_1^{\left( 4 \right)}$ & ${\left( {1:12:133 } \right)^{\rm{T}}}$\\
		5 & ${\left( 49:1:60 \right)^{\rm{T}}}$ & ${\bf{W}}_1^{\left( 5 \right)}$ & ${\left( {1:12:133 } \right)^{\rm{T}}}$\\
		6 & ${\left( 61:1:72 \right)^{\rm{T}}}$ & ${\bf{W}}_1^{\left( 6 \right)}$ & ${\left( {1:12:133 } \right)^{\rm{T}}}$\\
		7 & ${\left( 73:1:84 \right)^{\rm{T}}}$ & ${\bf{W}}_1^{\left( 6 \right)}$ & ${\left( {13:12:145 } \right)^{\rm{T}}}$\\
		8 & ${\left( 85:1:96 \right)^{\rm{T}}}$ & ${\bf{W}}_1^{\left( 6 \right)}$ & ${\left( {25:12:157 } \right)^{\rm{T}}}$\\
		$ \vdots $ & $ \vdots $ & $\vdots$ & $\vdots$\\
		94 & ${\left( 1117:1:1128 \right)^{\rm{T}}}$ & ${\bf{W}}_1^{\left( 6 \right)}$ & ${\left( {1057:12:1189 } \right)^{\rm{T}}}$\\
		95 & ${\left( 1129:1:1140 \right)^{\rm{T}}}$ & ${\bf{W}}_1^{\left( 7 \right)}$ & ${\left( {1057:12:1189 } \right)^{\rm{T}}}$\\
		96 & ${\left( 1141:1:1152 \right)^{\rm{T}}}$ & ${\bf{W}}_1^{\left( 8 \right)}$ & ${\left( {1057:12:1189 } \right)^{\rm{T}}}$\\
		97 & ${\left( 1153:1:1164 \right)^{\rm{T}}}$ & ${\bf{W}}_1^{\left( 9 \right)}$ & ${\left( {1057:12:1189 } \right)^{\rm{T}}}$\\
		98 & ${\left( 1165:1:1176 \right)^{\rm{T}}}$ & ${\bf{W}}_1^{\left( 10 \right)}$ & ${\left( {1057:12:1189 } \right)^{\rm{T}}}$\\
		99 & ${\left( 1177:1:1188 \right)^{\rm{T}}}$ & ${\bf{W}}_1^{\left( 11 \right)}$ & ${\left( {1057:12:1189 } \right)^{\rm{T}}}$\\
		100 & ${\left( 1190:1:1200 \right)^{\rm{T}}}$ & ${\bf{W}}_1^{\left( 12 \right)}$ & ${\left( {1057:12:1189 } \right)^{\rm{T}}}$\\
		\tabucline[2pt]{-}
	\end{tabu}
\end{table}

Several simplifications can be considered to simplify and to realize the design of LMMSE channel estimators. First, suppose the fading multipath channel in Formula \eqref{eq1} satisfies the following conditions. ${h\left( {{\tau _m}} \right)}$ are zero-mean complex random variables with a power-delay profile $\theta \left( {{\tau _m}} \right) = C{e^{ - {\tau _m}/{\tau _{rms}}}}$, and delays ${{\tau _m}}$ are uniformly and independently distributed. [4] indicates that the single element in the correlation matrix ${{\bf{R}}_{{\bf{H}}{{{\bf{\hat H}}}_{LS,r,s}}}}$ or ${{{\bf{R}}_{{{\bf{H}}_{r,s}}{{\bf{H}}_{r,s}}}}}$ between subcarrier ${k_a}$ and ${k_b}$ under this channel model is computed as follows:
\begin{equation}\label{eq16}
{r_{{k_a},{k_b}}} = \left\{ {\begin{array}{*{20}{l}}
	{\begin{array}{*{20}{c}}
		{}&{}&{1,}&{}
		\end{array}\begin{array}{*{20}{c}}
		{}&{\begin{array}{*{20}{c}}
			{}
			\end{array}\begin{array}{*{20}{c}}
			{}
			\end{array}\begin{array}{*{20}{c}}
			{}
			\end{array}if\begin{array}{*{20}{c}}
			{{k_a} = {k_b}}&{}
			\end{array}}
		\end{array}}\\
	{\begin{array}{*{20}{c}}
		{\frac{{1 - {e^{ - j2\pi L\frac{{{k_a} - {k_b}}}{N}}}}}{{j2\pi L\frac{{{k_a} - {k_b}}}{N}}},}&{\begin{array}{*{20}{c}}
			{}&{if\begin{array}{*{20}{c}}
				{{k_a} \ne {k_b}}&{}
				\end{array}}
			\end{array}}
		\end{array}}
	\end{array}} \right.
\end{equation}
where $L$ represents the assumptive number of channel paths, $N$ is the number of subcarriers, and ${k_a}$ and ${k_b}$ denote the indexes of subcarriers intended to calculate the correlation. Thus, each element in ${{\bf{R}}_{{\bf{H}}{{{\bf{\hat H}}}_{LS,r,s}}}}$ or ${{{\bf{R}}_{{{\bf{H}}_{r,s}}{{\bf{H}}_{r,s}}}}}$ only depends on the difference between ${k_a}$ and ${k_b}$ instead of the value of ${k_a}$ or ${k_b}$. Therefore, given $L$, $N$ and set $d = {k_a} - {k_b}$, Formula \eqref{eq16} transitions into the following:
\begin{equation}\label{eqrd}
{r_d} = \left\{ {\begin{array}{*{20}{l}}
	{\begin{array}{*{20}{c}}
		{}&{}&{1,}&{\begin{array}{*{20}{c}}
			{}&{\begin{array}{*{20}{c}}
				{}
				\end{array}if}
			\end{array}\begin{array}{*{20}{c}}
			{}
			\end{array}d = 0}
		\end{array}}\\
	{\begin{array}{*{20}{c}}
		{\frac{{1 - {e^{ - j2\pi L\frac{d}{N}}}}}{{j2\pi L\frac{d}{N}}},}&{if\begin{array}{*{20}{c}}
			{}
			\end{array}d \ne 0}
		\end{array}}
	\end{array}} \right.
\end{equation}
Robustness characteristics of the LMMSE estimator toward $L$ and ${\rm{SNR_{fixed}}}$ enable pre-calculation of the weighed matrix ${{\bf{W}}_{r,s}}$ under assumptive parameters $L$ and ${\rm{SNR_{fixed}}}$, which are present in the simulation results in Section \uppercase\expandafter{\romannumeral5}. Therefore, the weight matrix only depends on relative locations of the subcarrier under certain $L$ and ${\rm{SNR_{fixed}}}$, indicating that it forms a relationship with UE pilot locations and bears no relation with the receiving antennas. Thus, Formula \eqref{eq13} can be rewritten as follows:
\begin{equation}\label{eq17}
{{\bf{W}}_s} = {{\bf{R}}_{{{\bf{H}}_{r,s}}{{{\bf{\hat H}}}_{LS,r,s}}}}{\left( {{{\bf{R}}_{{{\bf{H}}_{r,s}}{{\bf{H}}_{r,s}}}} + \frac{{\sigma ^2}}{{{\sigma _s^2}}}{\bf{I}}} \right)^{ - 1}}.
\end{equation}

\subsubsection{ZF MIMO detector}:
The MIMO detector detects the original signal ${\bf{X}}$ of UE in Formula \eqref{eq7} from observation ${\bf{Y}}$ and estimates ${{\bf{\tilde X}}}$. The recovered data in ZF MIMO detector are calculated as follows:
\begin{equation}\label{eq18}
{{\bf{\tilde X}}} = {{\bf{W}}_{ZF}}{\bf{Y}},
\end{equation}
where
\begin{equation}\label{eq19}
{{\bf{W}}_{ZF}} = {\left( {{\bf{H}}^{\rm{H}}{{\bf{H}}}} \right)^{ - 1}}{\bf{H}}^{\rm{H}}.
\end{equation}

\subsubsection{MMSE MIMO detector}:
The MMSE MIMO detector can maximize signal-to-interference-plus-noise ratio after detection, and detection is similar to that in Formula \eqref{eq18}. 
\begin{equation}\label{eq20}
{{\bf{\tilde X}}} = {{\bf{W}}_{MMSE}}{\bf{Y}},
\end{equation}
where
\begin{equation}\label{eq21}
{{\bf{W}}_{MMSE}} = {\left( {{\bf{H}}^{\rm{H}}{{\bf{H}}} + \sigma ^2{\bf{I}}} \right)^{ - 1}}{\bf{H}}^{\rm{H}}.
\end{equation}

Formulas \eqref{eq19} and \eqref{eq21} indicate that the difference in ZF and MMSE MIMO detector is that the MMSE MIMO detector must know the prior noise variance. However, both detectors must still calculate the matrix inverse, which is time-consuming.

Thus, to address this problem, Formulas \eqref{eq20} and \eqref{eq21} are reconsidered to find ${\bf{\tilde X}}$, which is equivalent to solving the linear function:
\begin{equation}\label{eq22}
\left( {{{\bf{H}}^{\rm{H}}}{\bf{H}} + {\sigma ^2}{\bf{I}}} \right){\bf{\tilde X}} = {{\bf{H}}^{\rm{H}}}{\bf{Y}},
\end{equation}
which  will be utilized in convenient implementation of RaPro.

\subsection{3W-LMMSE and 12W-LMMSE Mapping Rules Design}
The BS in the massive MIMO system is equipped with an increasing number of antennas to gain improved accuracy performance, causing the rising complexity in channel estimators. Formula \eqref{eq17} shows that the matrix inverse operation dominates computational complexity of LMMSE channel estimation. The matrix dimension must be reduced to obtain a low-complexity LMMSE channel estimator. According to dimension reduction, we design two mapping rules [5], 3W-LMMSE and 12W-LMMSE, which reduce the matrix dimension to 3$\times$3 and 12$\times$12, respectively.

\subsubsection{3W-LMMSE channel estimator}:
The complexity of LMMSE channel estimate calculation according to Formula \eqref{eq17} mainly draws upon the inverse of a matrix. The matrix features a 1200$\times$1200 dimension in the presence of 1200 subcarriers. The 3W-LMMSE channel estimator can reduce this computation effectively by separating a 1200$\times$1200 dimension matrix inverse into 100 times 3$\times$3 dimension matrix inverse with some performance trade-off. The 3W-LMMSE channel estimator partitions the $K$ LS estimate into $K$ groups, and each group contains three elements through the overlap. Each LS-group-constituted vector is multiplied by a 3$\times$3-dimension weight matrix generated from Formula \eqref{eqrd}. Then, $K$ group results are sequenced. The obtained 3$K$ channel estimate is the result of 3W-LMMSE estimator. Three weight matrices are needed in this process instead of 100 because of the simplified channel model. Thus, this method is named 3W-LMMSE.

We illustrate the idea by assuming a system with ${N = 1200}$ subcarriers in the resource block, ${R = 16}$ receiving antennas, and ${S = 12}$ UEs, in which only four UEs are active. Based on the assumed pilot structure, the groups of UE 1 is tabulated in Table \uppercase\expandafter{\romannumeral1}. The table shows that the difference between each element in the subcarriers of ${{{\bf{\hat H}}}_{LMMSE,r,s}}$ and the subcarriers of ${{{\bf{\hat H}}}_{LS,r,s}}$ is equivalent from Indexes 2 to 99; thus, ${{\bf{R}}_{{\bf{H}}{{{\bf{\hat H}}}_{LS,r,s}}}}$ is the same. The difference between elements in subcarriers inside ${{{\bf{\hat H}}}_{LS,r,s}}$ is equivalent from Indexes 1 to 100; thus, ${{{\bf{R}}_{{{\bf{H}}_{r,s}}{{\bf{H}}_{r,s}}}}}$ is the same. Formula \eqref{eqrd} shows that weight matrix is the same from Indexes 2 to 99: ${{\bf{W}}^{\left( 2 \right)}}$.

The weight matrix for other UEs must also be considered. Variations in pilot location result in the same effect on the subcarriers of ${{{\bf{\hat H}}}_{LMMSE,r,s}}$ and ${{{\bf{\hat H}}}_{LS,r,s}}$. Thus, ${{\bf{R}}_{{\bf{H}}{{{\bf{\hat H}}}_{LS,r,s}}}}$ remains unchanged. Similarly, ${{\bf{R}}_{{\bf{H}}{{{\bf{\hat H}}}_{LS,r,s}}}}$ is also unchanged. Thus, in Table \uppercase\expandafter{\romannumeral1}, ${{\bf{W}}^{\left( 1 \right)}}$, ${{\bf{W}}^{\left( 2 \right)}}$, and ${{\bf{W}}^{\left( 3 \right)}}$ correspond to weight matrices for all UEs and their values are as follows:\\
\begin{equation}\label{eqW1}
{{\bf{W}}^{\left( 1 \right)}} = \left[ {\begin{array}{*{20}{c}}
	{{r_0}}&{{r_{ - 12}}}&{{r_{ - 24}}}\\
	{{r_4}}&{{r_{ - 8}}}&{{r_{ - 20}}}\\
	{{r_8}}&{{r_{ - 4}}}&{{r_{ - 16}}}
	\end{array}} \right]{\bf{P}},\\
\end{equation}

\begin{equation}\label{eqW2}
{{\bf{W}}^{\left( 2 \right)}} = \left[ {\begin{array}{*{20}{c}}
	{{r_{12}}}&{{r_0}}&{{r_{ - 12}}}\\
	{{r_{16}}}&{{r_4}}&{{r_{ - 8}}}\\
	{{r_{20}}}&{{r_8}}&{{r_{ - 4}}}
	\end{array}} \right]{\bf{P}},\\
\end{equation}

\begin{equation}\label{eqW3}
{{\bf{W}}^{\left( 3 \right)}} = \left[ {\begin{array}{*{20}{c}}
	{{r_{24}}}&{{r_{12}}}&{{r_0}}\\
	{{r_{28}}}&{{r_{16}}}&{{r_4}}\\
	{{r_{32}}}&{{r_{20}}}&{{r_8}}
	\end{array}} \right]{\bf{P}},
\end{equation}
where

\begin{equation}\label{eqP}
{\bf{P}} = \left( {\left[ {\begin{array}{*{20}{c}}
		{{r_0}}&{{r_{ - 12}}}&{{r_{ - 24}}}\\
		{{r_{12}}}&{{r_0}}&{{r_{ - 12}}}\\
		{{r_{24}}}&{{r_{12}}}&{{r_0}}
		\end{array}} \right] + \frac{{{\sigma ^2}}}{{\sigma _s^2}}{{\bf{I}}_3}} \right).\\
\end{equation}

\begin{figure}[!t]
	\centering
	\includegraphics[width=3in]{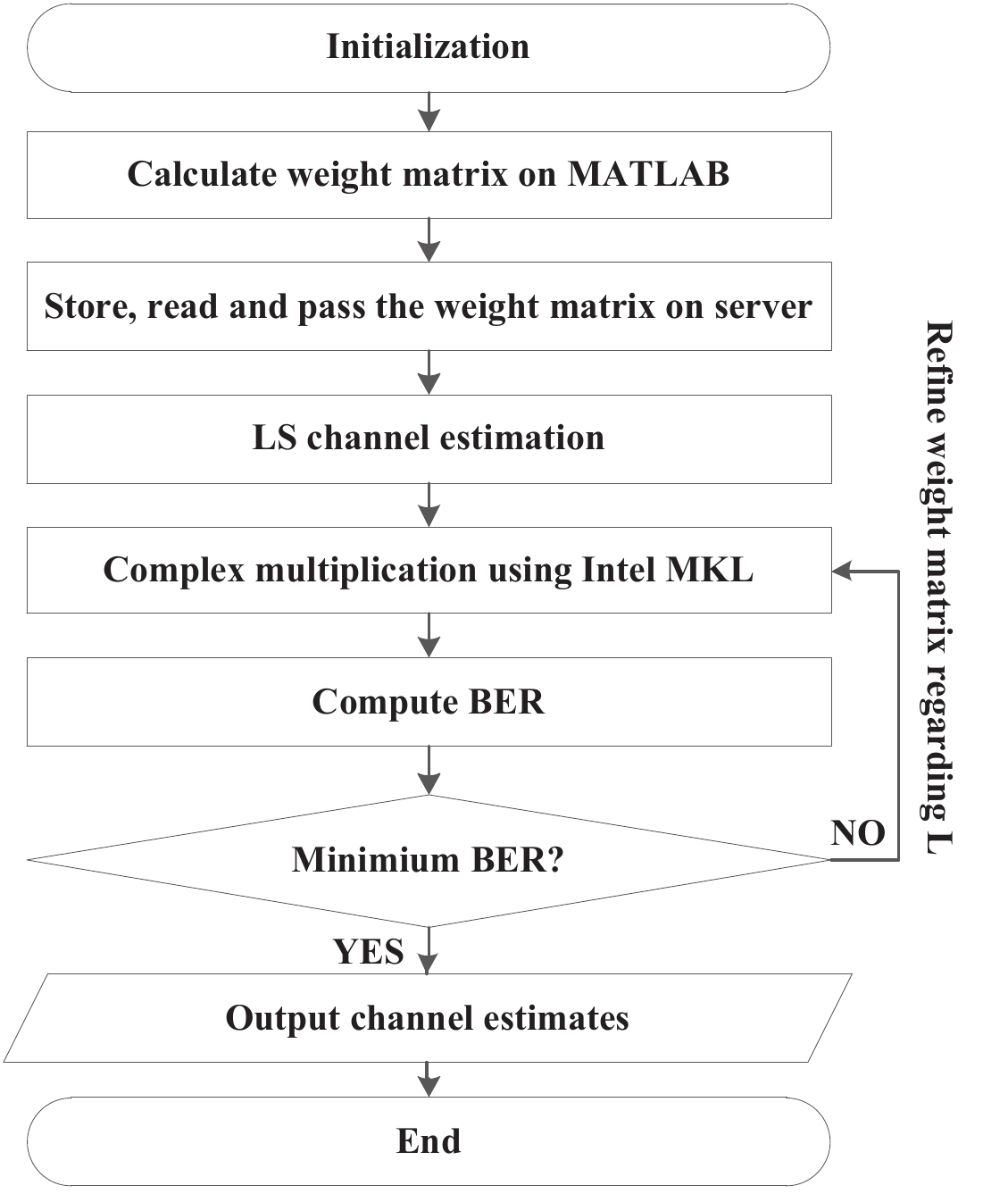}
	\caption{Implementation procedures of low-complexity LMMSE channel estimation. Weight matrices should be first calculated on MATLAB and passed to the function of channel estimation on a server as parameters. Combined with receiving data and local pilots, LS channel estimates can be obtained, and low-complexity LMMSE, 3W-LMMSE, or 12W-LMMSE estimates are included in the complex multiplication of weight matrix and LS channel estimates. $L$ value can be refined based on BER performance to fit the test environment.}
	\label{fig5}
\end{figure}
\begin{figure}[!t]
	\centering
	\includegraphics[width=3.5in]{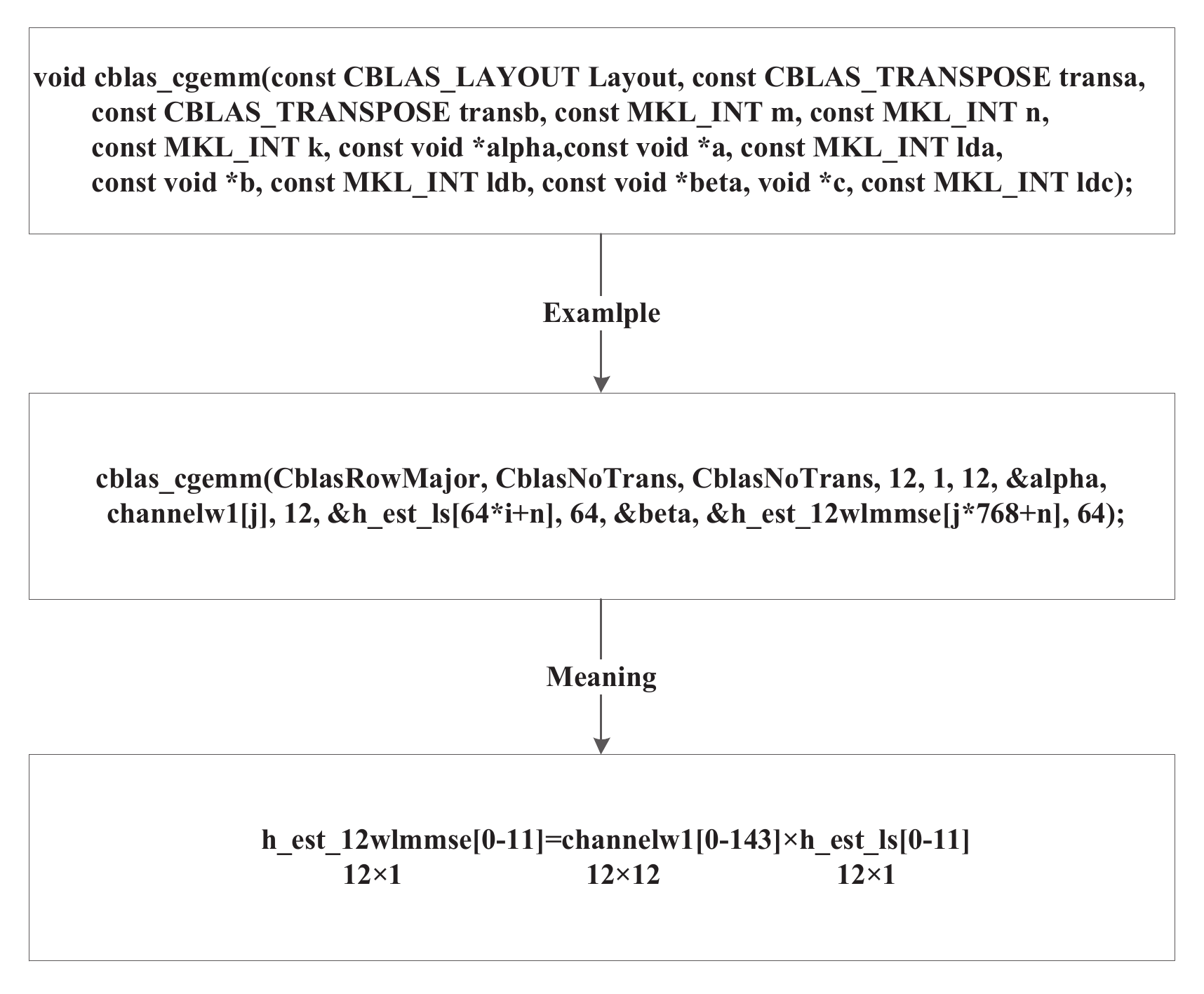}
	\caption{Intel MKL interpretation of complex multiplication. The function ``cblas\_cgemm'' is used for complex multiplication of matrices or vectors. A typical example is shown in the middle box, whose meaning is explained in the bottom box, which presents that the function calculates multiplication of 12$\times$12-dimension matrix and 12$\times$1 vector.}
	\label{fig6}
\end{figure}

\subsubsection{12W-LMMSE channel estimator}:
Compared with 3W-LMMSE, a 12W-LMMSE channel estimator employs 12 subcarriers to calculate the correlation matrix. Thus, the complexity of 12W-LMMSE is more complex than that of 3W-LMMSE but still less than that of a LMMSE channel estimator. The 12W-LMMSE group features the same goal as the 3W-LMMSE group, whereas 12 elements are identified in each vector, and the weight matrix presents a 12$ \times $12 dimension. Table \uppercase\expandafter{\romannumeral2} provides data on the 12W-LMMSE group, where ${\left( {1:1:12 } \right)^{\rm{T}}}$ is interpreted as 1,2,...,12, and ${\left( {1:12:133 } \right)^{\rm{T}}}$ is interpreted as 1,13,...,133. LMMSE estimate can be calculated by multiplication of the weight matrix and the corresponding LS estimate vector.

The 12W-LMMSE estimator is highly complicated considering its implementation. In addition to the 100 computations of 12$ \times $12 matrix inverse, 100 times $ \left( {12 \times 12} \right) \times \left( {12 \times 1} \right) $ complex multiplication is needed with 12 weight matrices. On the other hand, only three weight matrices are needed in 3W-LMMSE to conduct 100 times $\left( {3 \times 3} \right) \times \left( {3 \times 1} \right)$ complex multiplication. Different UEs in 12W-LMMSE lead to different subcarriers of ${{{\bf{\hat H}}}_{LS,r,s}}$. Although the subcarriers of ${{{\bf{\hat H}}}_{LMMSE,r,s}}$ show no changes, ${{\bf{R}}_{{\bf{H}}{{{\bf{\hat H}}}_{LS,r,s}}}}$ still changes from UE to UE, which causes different UEs to feature different weight matrix groups. That is, each UE contains 12 weight matrices that differ from those of others, namely, ${\bf{W}}_s^{\left( 1 \right)}$ to ${\bf{W}}_s^{\left( 12 \right)}$, where $1 \le s \le 4$.

\begin{figure*}[!t]
	\centering
	\subfloat[ ]{
		\includegraphics[width=.5\textwidth]{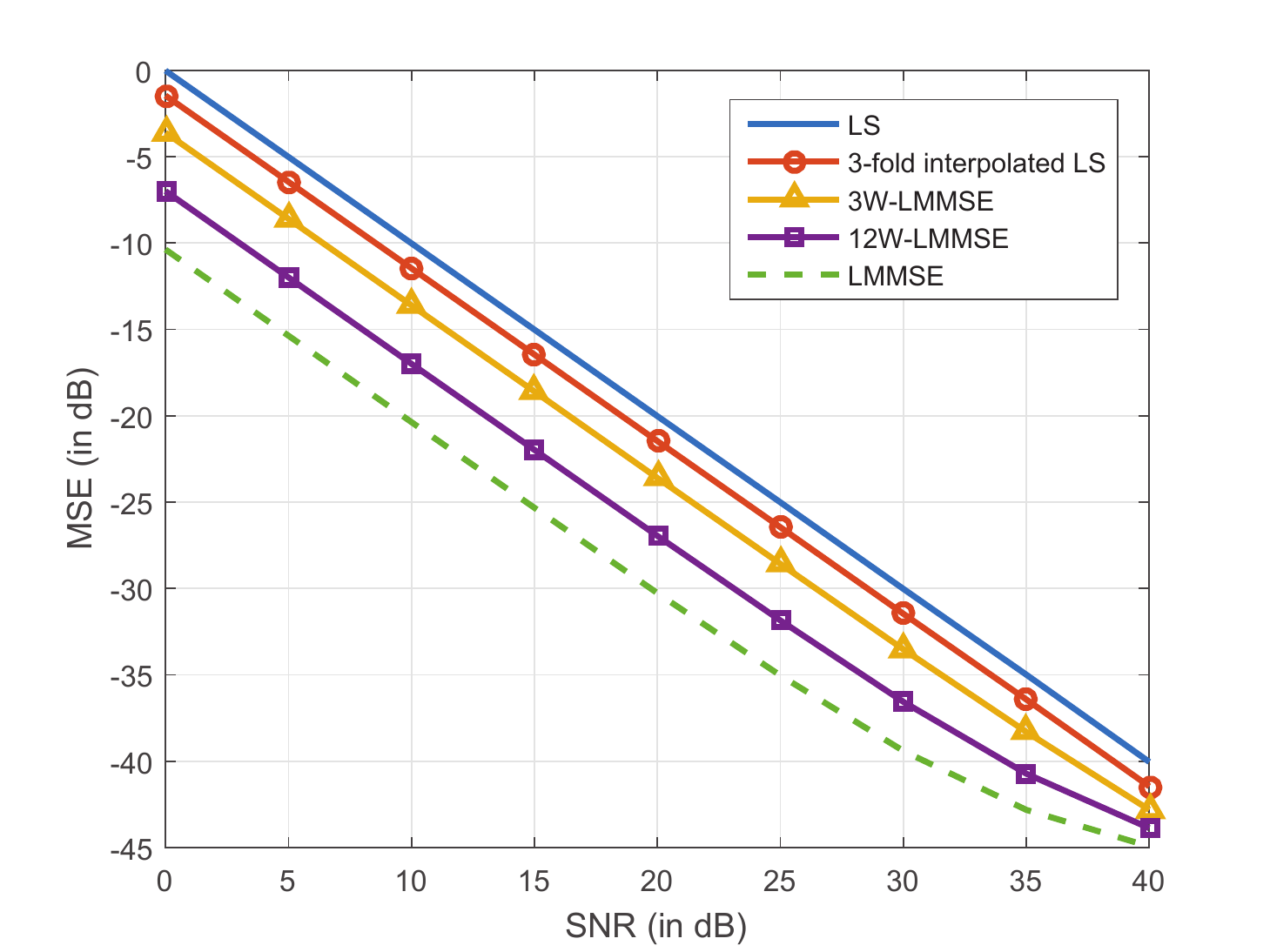}}\hfill
	\subfloat[ ]{
		\includegraphics[width=.5\textwidth]{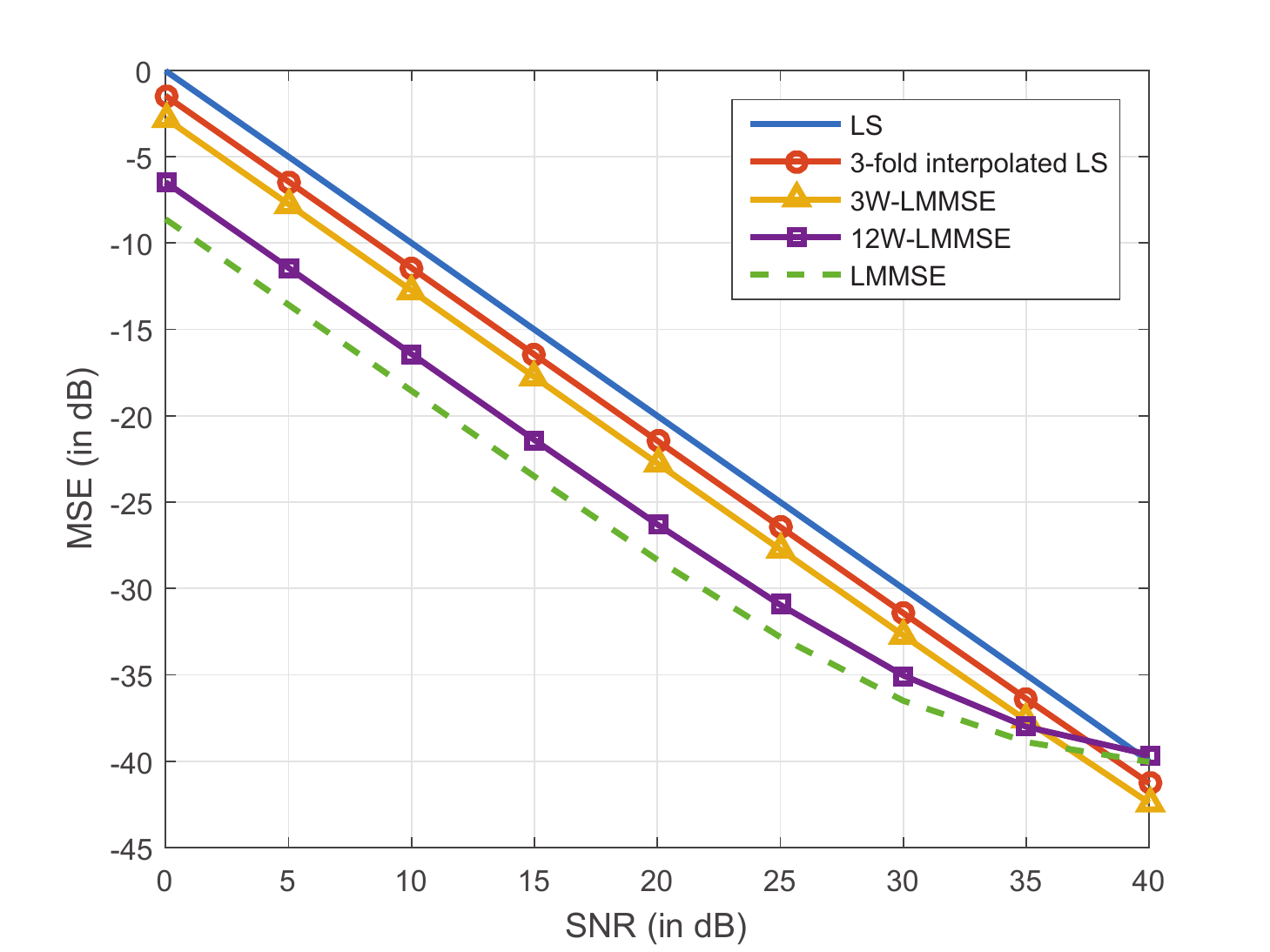}}\\
	\subfloat[ ]{
		\includegraphics[width=.5\textwidth]{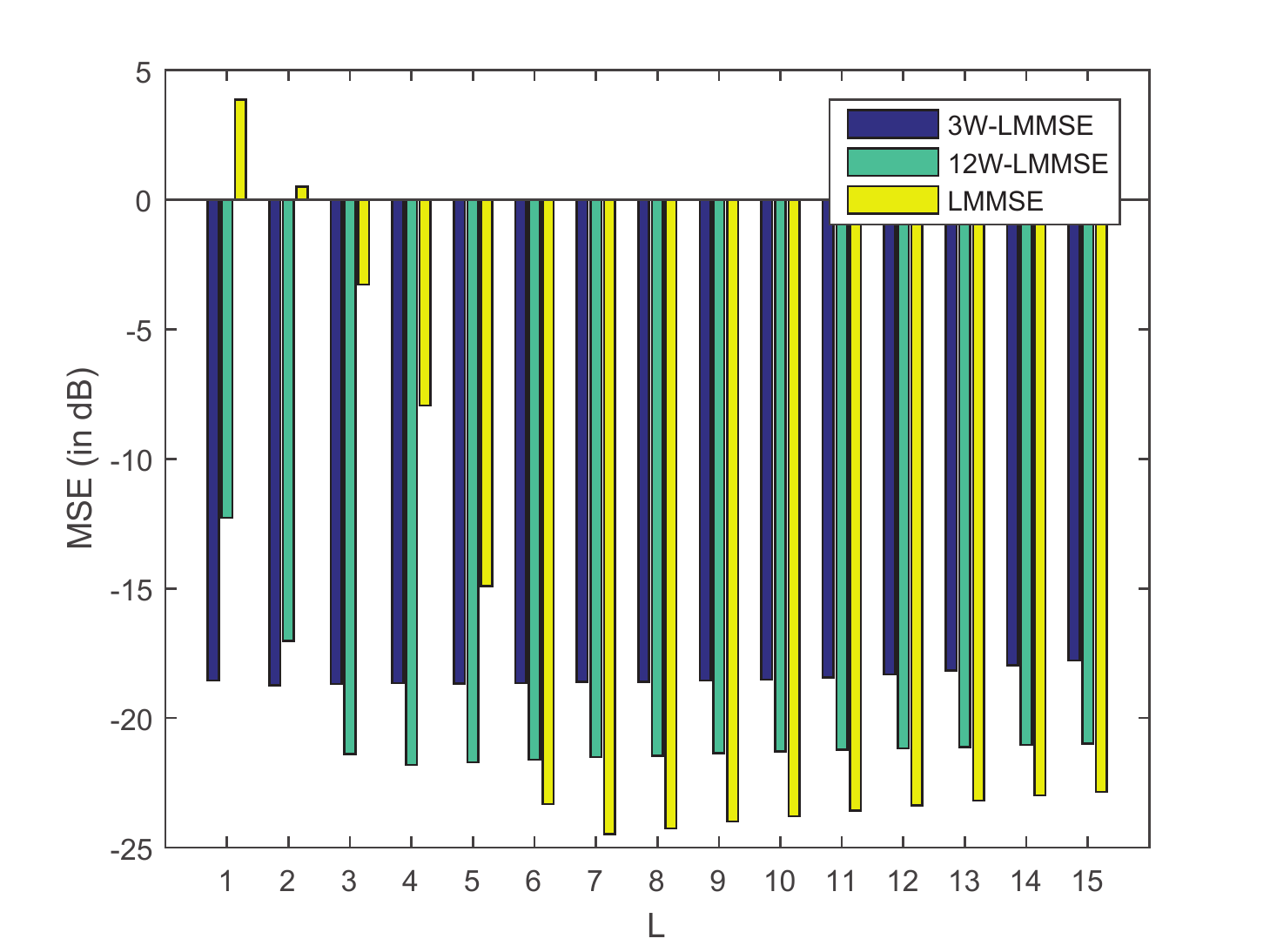}}\hfill
	\subfloat[ ]{
		\includegraphics[width=.5\textwidth]{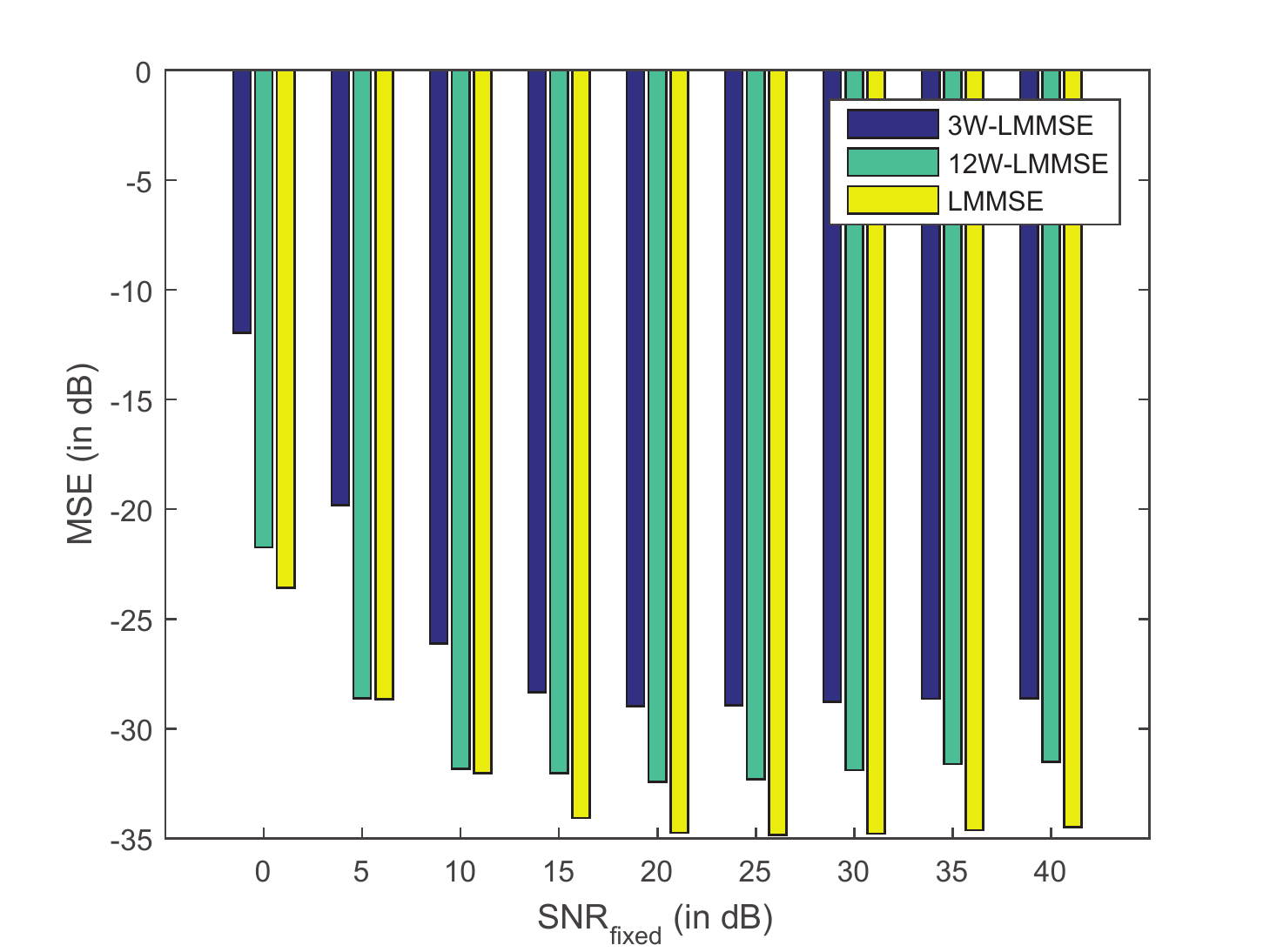}}
	\caption{Simulation results for channel estimation schemes: (a) MSE performance of channel estimation schemes without timing errors. (b) MSE performance of channel estimation schemes with a two-point timing error. (c) MSE performance changes of MMSE-based channel estimation schemes along with $L$, which varies from 1 to 15 when the realistic value is 7. (d) MSE performance changes of MMSE-based channel estimation schemes along with $ {\rm{SNR_{fixed}}} $ varying from 0 dB to 40 dB when the realistic value is 25 dB.}
	\label{fig7}
\end{figure*}

\subsection{Detailed Implementation Procedures on RaPro}
Implementation procedures of channel estimators and MIMO detector are packaged as functions invoked by the slot processing threads (Figure \ref{fig4}). This section presents detailed implementation procedures within ``channel\_estimation.c'' and ``MMSE\_MIMO\_detection.c.''

\subsubsection{channel\_estimation.c}
Figure \ref{fig5} shows the detailed 12W-LMMSE channel estimation procedure. First, the weight matrix ${\bf{W}}_s^{\left( 1 \right)}$ to ${\bf{W}}_s^{\left( 12 \right)}$ for each UE is calculated based on Formula \eqref{eq17} and Table \uppercase\expandafter{\romannumeral2} under fixed $ {\rm{SNR_{fixed}} = 40 dB} $ and different $L$ values, which vary from 1 to 15 in MATLAB. For one UE with a certain $L$ value, the weight matrix group includes 12 weight matrices, and each weight matrix contains 144 elements. Considering hardware implementation, the real and imaginary parts of these 12 weight matrices are written in two .txt files with 12$ \times $144 elements in one row. These parts are combined as a float complex array, as shown in the eclipse in Figure \ref{fig4}. 
\begin{figure*}[!t]
	\centering
	\subfloat[ ]{
		\includegraphics[width=.5\textwidth]{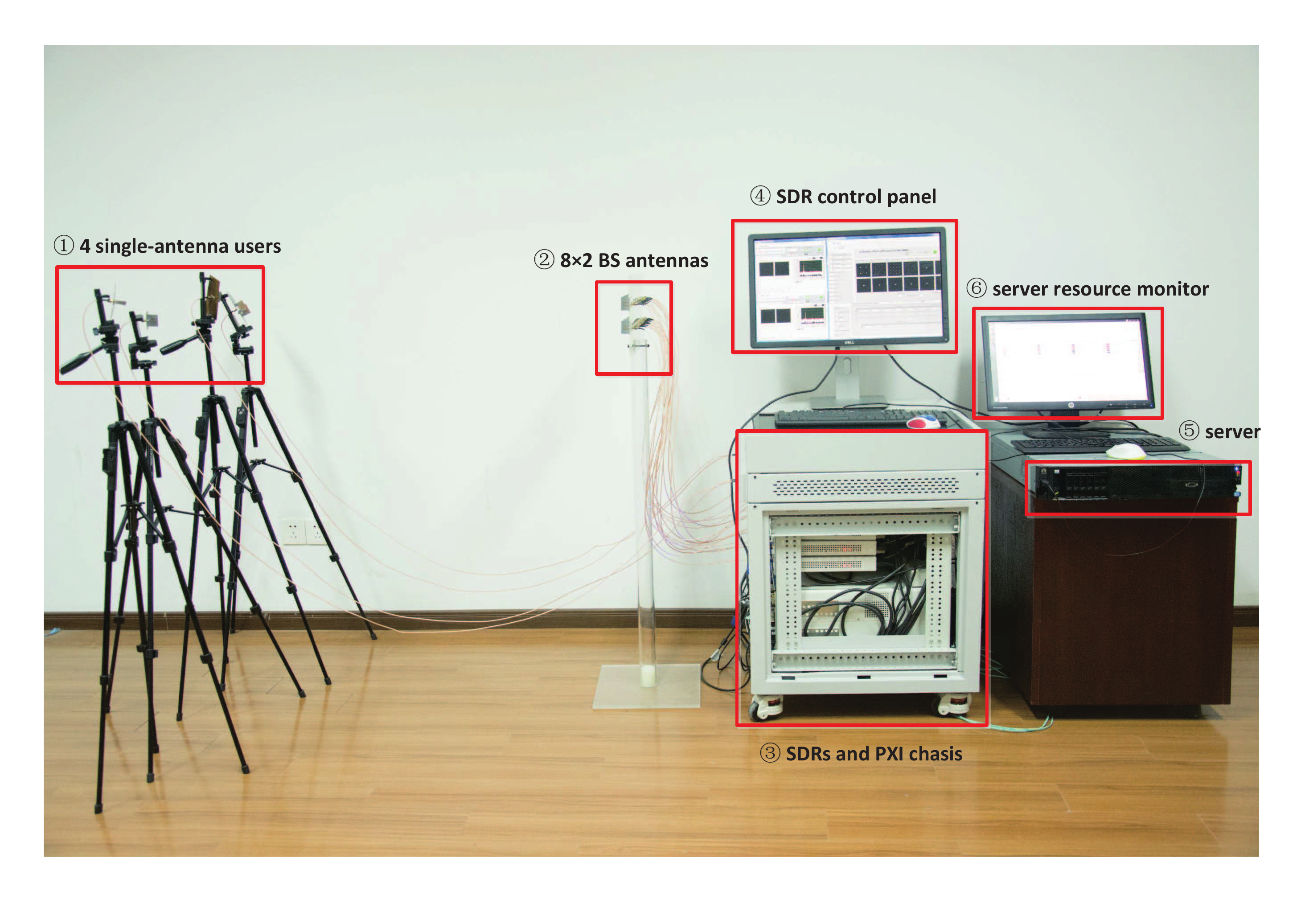}}\hfill
	\subfloat[ ]{
		\includegraphics[width=.5\textwidth]{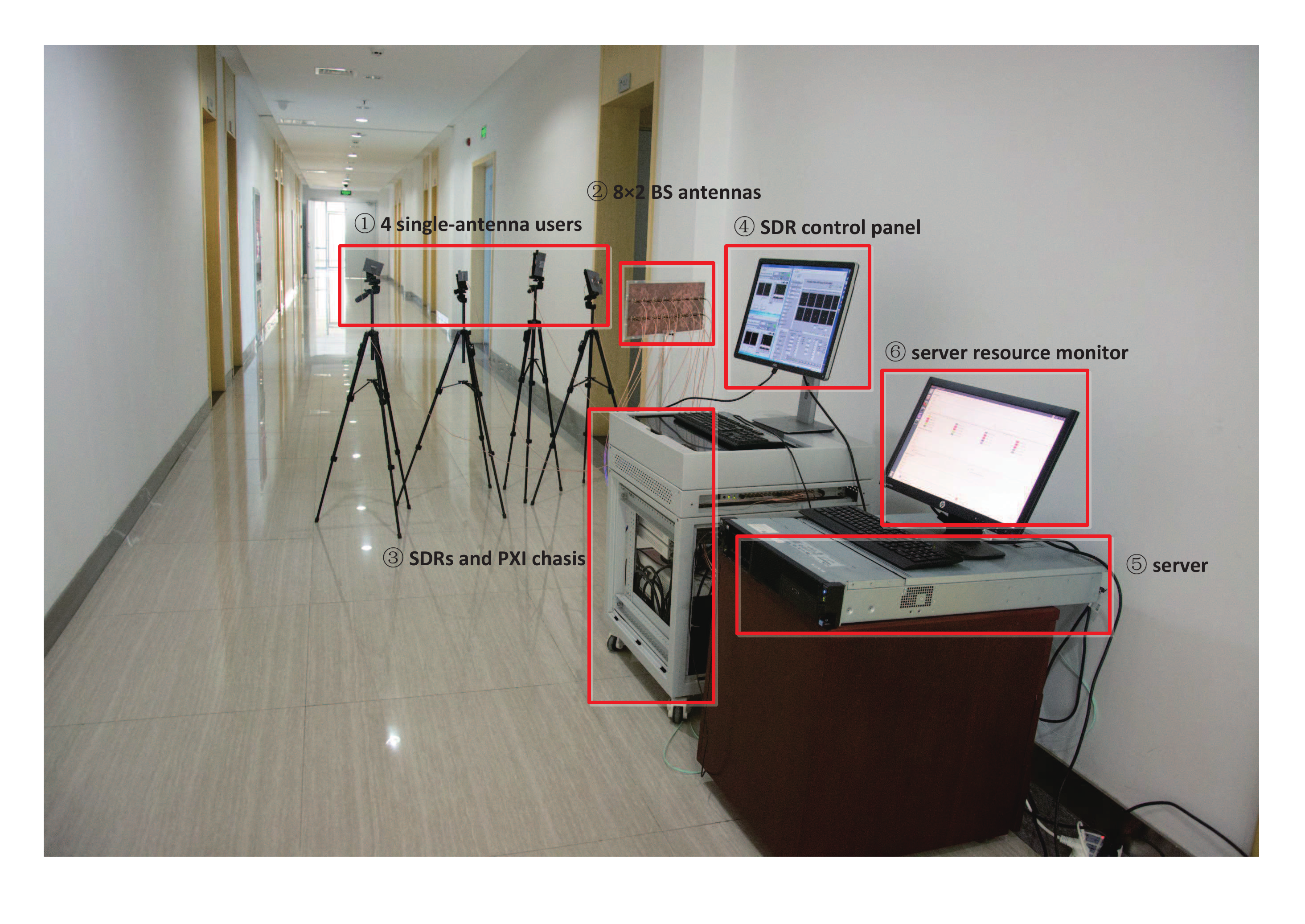}}\\
	\subfloat[ ]{
		\includegraphics[width=.97\textwidth]{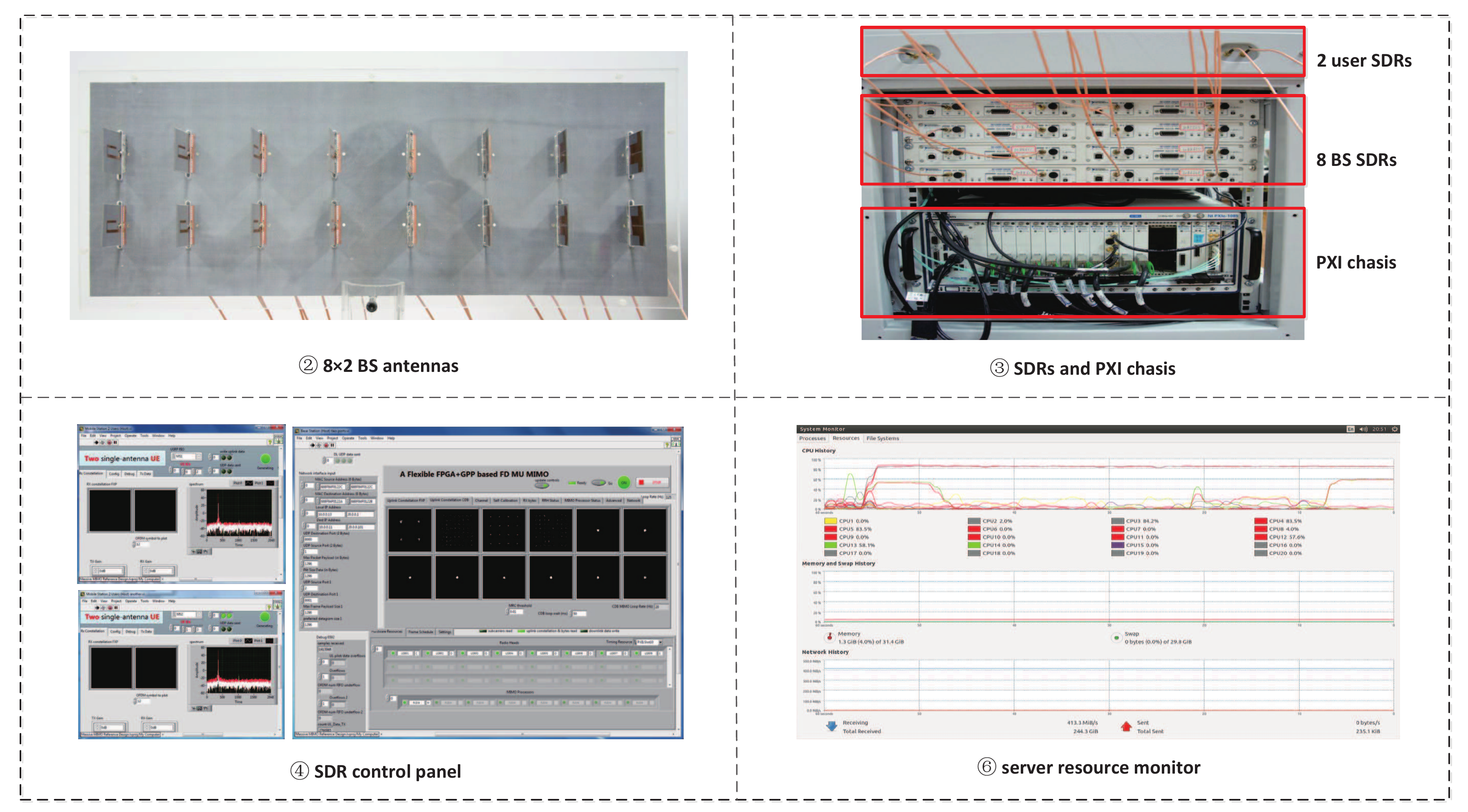}}
	\caption{Test scenarios and deployment of RaPro. (a) Indoor scenario occurs in a typical office room with an area of approximately 30 m$^2$. (b) Corridor scenario transpires in a 25 m long narrow corridor. (c) Details of some components include those of 8$\times$2 BS antennas, two UE SDRs, eight BS SDRs, a PXI chassis, a SDR control panel, and a server resource monitor.}
	\label{fig8}
\end{figure*}

Next, 15 $\times$ 4 $\times$ 2 (15 $L$ values, 4 UEs, and 2 for real part and image part) .txt files are stored on the server and read into parameter W in C file before starting the main thread. Parameter W contains information of the weight matrix and is passed to the slot processing thread, in which the 12W-LMMSE channel estimation algorithm within one slot is deployed. Other data needed to conduct channel estimation include the received UDP data and local pilots. Thus, these data are also passed to this thread.

Deployment of 12W-LMMSE channel estimation can be divided into two steps. The first step calculates the LS channel estimation by multiplying data in the pilot location of the first OFDM symbol by the local pilot according to Formula \eqref{eq9}. In this paper, we replace division with multiplication as the magnitude of quadrature phase-shift keying (QPSK)-modulated pilot symbol is normalized to 1, and the local pilot is a conjugated pilot. The next step performs the complex multiplication of the weight matrix and LS channel estimation results obtained from the former step to obtain $ {{{\bf{\hat H}}}_{LMMSE,r,s}} $ based on Formula \eqref{eq12}.

Some attention must focus on related engineering implementation. As previously analyzed, each UE possesses its own weight matrix group in the 12W-LMMSE channel estimator, which differs from the 3W-LMMSE channel estimator that shares the same weight matrix group among UEs. Thus, data belonging to different UEs must be distinguished and multiplied by a corresponding weight matrix in a 12W-LMMSE channel estimator.

Considering robustness simulation results regarding parameter $L$, the most adequate $L$ can be found by selecting the weight matrix group with the largest $L$, checking BER performance. BER performance is then continually checked using the weight matrix comprising a group with one less $L$ and another group until the best performance is achieved. This group is regarded as the most suitable weight matrix under the current scenario.

Another point to consider is the means of conducting complex multiplication between a matrix, such as ${\bf{W}}_s^{\left( 1 \right)}$, and a vector, such as a subcarrier group of $ {{{\bf{\hat H}}}_{LS,r,s}} $. The solution employs a function in Intel MKL to simplify and accelerate calculation instead of using several loops to calculate scalar elements individually and arranging them in a sequence.

For each UE, 100 times of complex multiplication are performed in 12W-LMMSE channel estimation, implying the importance of Intel MKL. Intel MKL provides math routines and functions with improved performance for software applications that solve large computational problems. The program is optimized for the latest Intel processors, which include those with multiple cores. Routines for BLAS Level 1, 2, and 3 in Intel MKL are designed for vector-vector, matrix-vector, and matrix-matrix operations, respectively. Matrix-matrix operations are used for generalized purposes.

Intel MKL indicates that the function ``cblas\_cgemm'' computes a complex matrix product with general matrices. The syntax of function ``cblas\_cgemm'' is shown as the top box in Figure \ref{fig6}. The operation of this function is defined as ${C: = alpha \times op(A) \times op(B) + beta \times C}$, where $op(X)$ is either $X, {X^T},$ or ${X^H}$; $alpha$ and $beta$ are scalars; $A$, $B$ and $C$ are matrices. $op(A)$ is an  $m$-by-$k$ matrix, $op(B)$ is a $k$-by-$n$ matrix, $C$ is an $m$-by-$n$ matrix. The input parameters imply the state of factors. $Layout$ specifies whether a two-dimensional array storage is row-major (CblasRowMajor) or column-major (CblasColMajor). For example, CblasRowMajor means that the matrix gradually stretches into a vector row by row. $transa$ specifies the form of $op(A)$, which is used in multiplication: $A$ (CblasNoTrans), $A^T$ (CblasTrans) or $A^H$ (CblasConjTrans). $transb$ is similar to $transa$, but it is used to specify $op(B)$. $a$ and $b$ refer to the initial positions of complex matrix factors, and $c$ is the initial position of the result. $lda$, $ldb$ and $ldc$ indicate the number of row or column in Matrices $A$, $B$ or $C$, respectively, according to $Layout$ [9]. 

A typical example of the complex multiplication between matrix ${\bf{W}}_1^{\left( 1 \right)}$ and the subcarrier group of $ {{{\bf{\hat H}}}_{LS,r,1}} $ vector is presented in the middle box of Figure \ref{fig6}. In this example, initialization values $alpha = 1$, $beta = 0$, $n = 0$ denote the transmission between UE 1 and receiver antenna 1, $i = 0$ denotes the first group of LS subcarriers, and $ j = 0$ denotes the first group of LMMSE subcarriers. Therefore, the operation calculates the content in the bottom box of Figure \ref{fig6}.

\subsubsection{MMSE\_MIMO\_detection.c}
In RaPro, MMSE MIMO detector is adopted to recover raw UE data according to Formula \eqref{eq22}. MIMO detection module is packed into a function ``MMSE\_MIMO\_detection.c.'' This section describes in detail the realization procedure of this function.

\begin{table}[H]
	\renewcommand{\arraystretch}{1.3}
	\caption{System Simulation Parameters}
	\label{tabel3}
	\centering
	\begin{tabu}{lll}
		\tabucline[2pt]{-}
		\textbf{Parameters} & \textbf{Variable} & \textbf{Value} \\
		\hline
		\# of BS antennas & $R$ & 16 \\
		\# of single-antenna UEs & $S$ & 4 \\
		FFT size & $N_{FFT}$ & 2048 \\
		\# of used subcarriers& $N$ & 1200\\
		CP type & - & Normal\\
		Modulation & - & 64-QAM\\
		Channel Type & - & Multipath channel\\
		\# of channel taps & $M$ & 6\\
		Channel delay & $\tau_m$ & [0,1,2,3,4,5]\\
		Channel tap power profile (dB)& - & [-2,-8,-10,-12,-15,-18]\\
		Assumptive SNR & ${\rm{SNR_{fixed}}}$ & 40dB\\
		Assumptive \# of channel taps & $L$ & 8\\
		\tabucline[2pt]{-}
	\end{tabu}
\end{table}

\begin{table*}
	\renewcommand{\arraystretch}{1.3}
	\caption{The BER values of 3W-LMMSE and 12W-LMMSE channel estimators with varing $L$ in scenarios: \ding{172} indoor 3W-LMMSE, \ding{173} indoor 12W-LMMSE, \ding{174} corridor 3W-LMMSE, \ding{175} corridor 12W-LMMSE. (The BER value equals to the value in table multiplies by ${10^{ - 6}}$)}
	\centering
	\begin{tabular}{p{0.7cm}<{\centering}p{0.7cm}<{\centering}p{0.7cm}<{\centering}p{0.7cm}<{\centering}p{0.7cm}<{\centering}p{0.7cm}<{\centering}p{0.7cm}<{\centering}p{0.7cm}<{\centering}p{0.7cm}<{\centering}p{0.7cm}<{\centering}p{0.7cm}<{\centering}p{0.7cm}<{\centering}p{0.7cm}<{\centering}p{0.7cm}<{\centering}p{0.7cm}<{\centering}p{0.7cm}<{\centering}}	
		\toprule[1.2pt]
		\textbf{$L$} & \textbf{1} &  \textbf{2} & \textbf{3} & \textbf{4} & \textbf{5} & \textbf{6} & \textbf{7} & \textbf{8} & \textbf{9} & \textbf{10} & \textbf{11} & \textbf{12} & \textbf{13} & \textbf{14} & \textbf{15}\\
		\hline		
		\ding{172} & 18 & 76  & 58 &  64 & 57 & 64 & 78 & 82 & 99 & 125 & 152 & 190 & 238 & 265 & 309\\
		\ding{173}  & 2 & 4 & 9 & 12 & 12 & 9 & 8 & 8 & 12 & 15 & 17 & 19 & 23 & 28 & 30\\
		\hline
		\ding{174} & 24 & 40 & 48 & 51 & 64 & 57 & 61 & 61 & 76 & 79 & 85 & 89 & 107 & 124 & 153\\
		\ding{175} & 0 & 0 & 0 & 0 & 0 & 0 & 0 & 1 & 1 & 1 & 1 & 1 & 1 & 2 & 3\\
		\bottomrule[1.2pt]
		\hline
	\end{tabular}
\end{table*}

\begin{figure*}[!t]
	\centering
	\subfloat[ ]{
		\includegraphics[width=.5\textwidth]{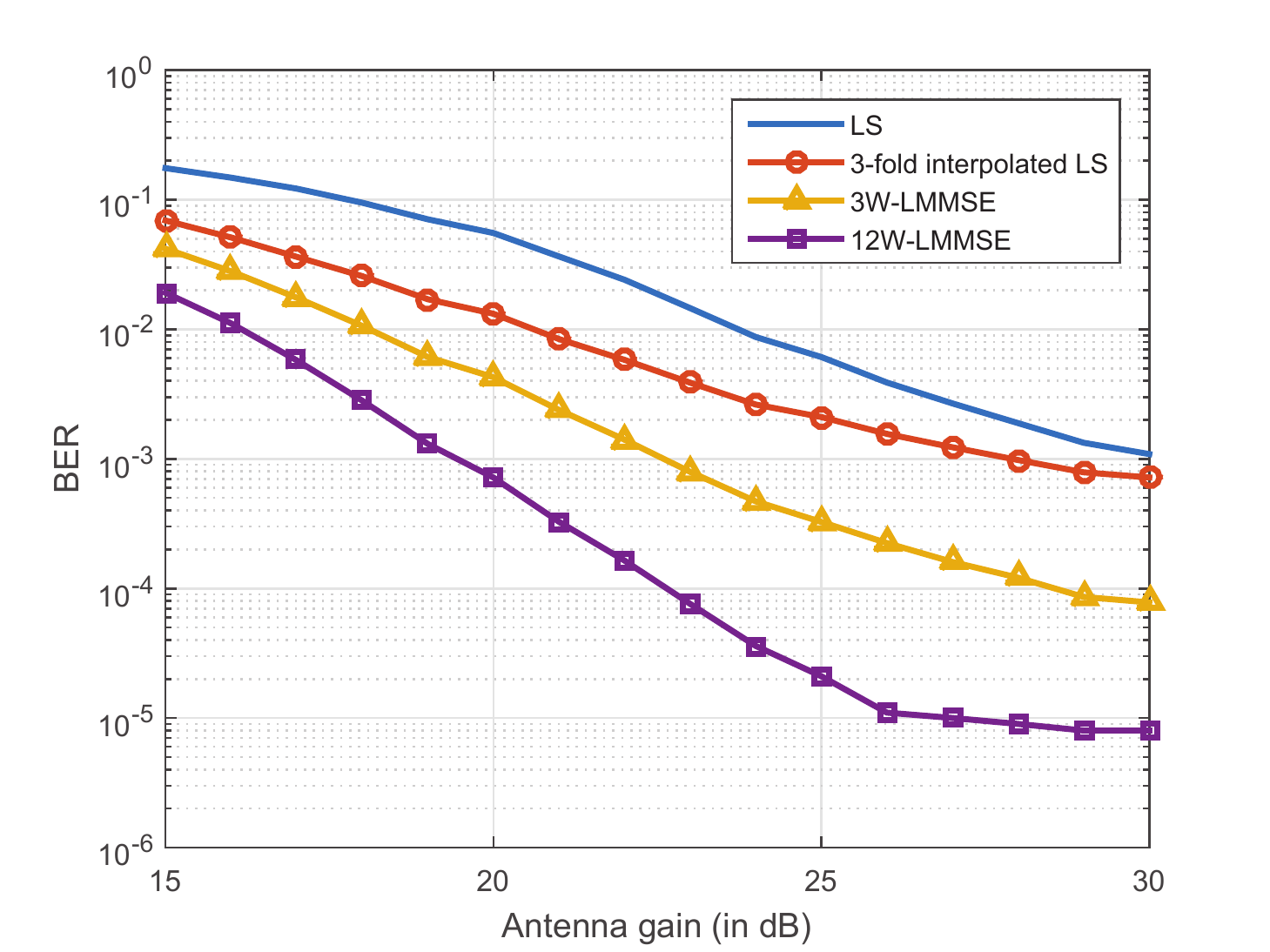}}\hfill
	\subfloat[ ]{
		\includegraphics[width=.5\textwidth]{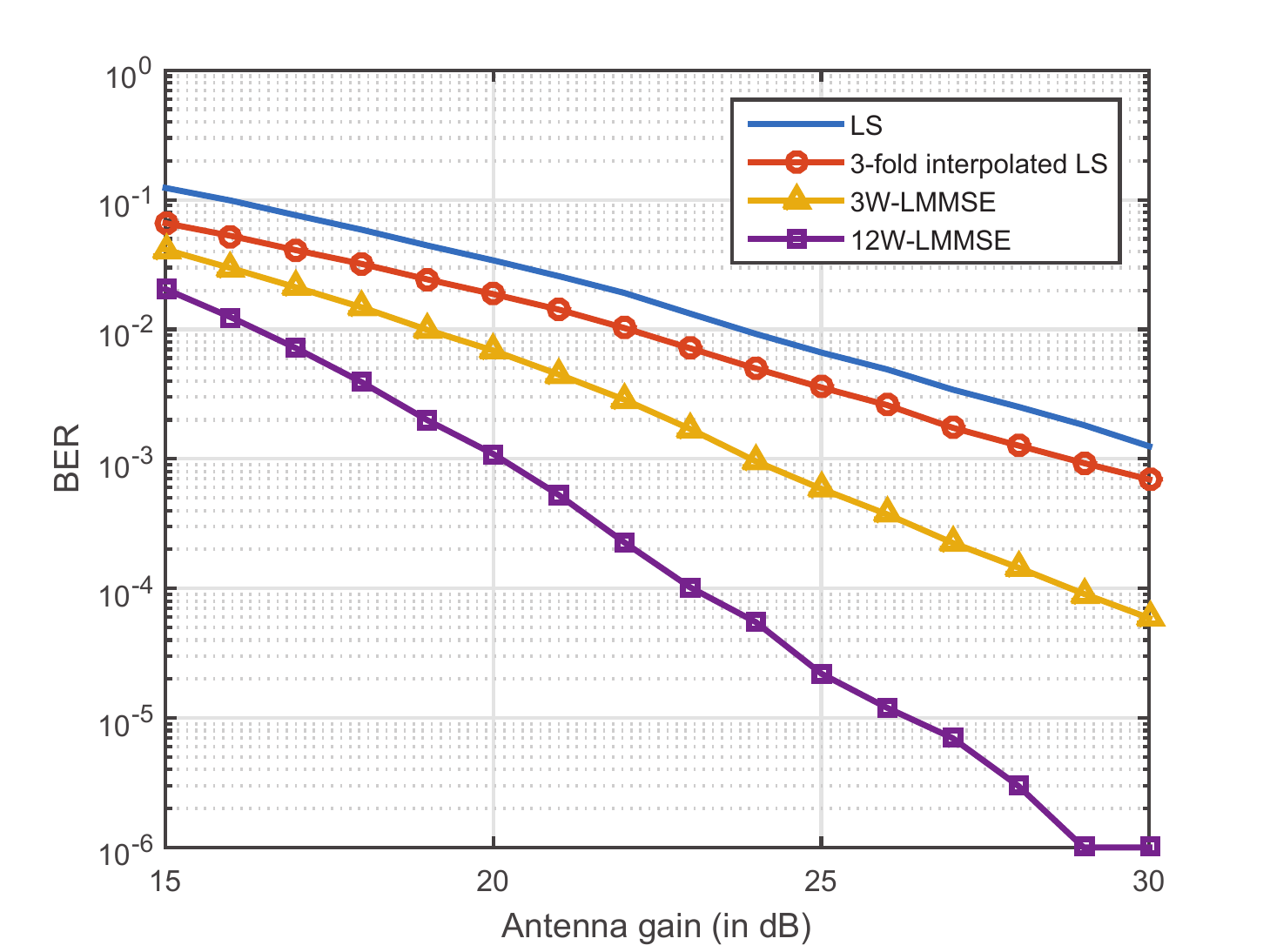}}
	\caption{Scenario test results for channel estimation schemes. (a) BER performances of different channel estimation schemes in an indoor test. (b) BER performances of different channel estimation schemes in a corridor test.}
	\label{fig9}
\end{figure*}

The first step involves preparation of prior data, which include the channel matrix $\bf{H}$ and noise variance ${{\sigma ^2} = 1000}$. Fixed ${{\sigma ^2} = 1000}$ is temporarily used for the lack of SNR estimation module. The channel matrix of each subcarrier comprises the element picked from the channel estimate passed from the ``channel\_estimation.c'' function. UEs data account for all the subcarriers, as shown in Figure \ref{fig1}. Thus, channel estimation should cover all the subcarriers to conduct detection on each. This situation naturally occurs for 12W-LMMSE channel estimation. For the other schemes mentioned, channel estimation results only consider upsampled subcarriers and must be complemented in advance. One of the most common idea is zero-order holding, which indicates that the following subcarrier blanks are padded with the last channel estimation value.

The components in Formula \eqref{eq22}, which include ${\left( {{{\bf{H}}^{\rm{H}}}{\bf{H}} + {\sigma ^2}{\bf{I}}} \right)}$ and ${{\bf{H}}^{\rm{H}}}{\bf{Y}}$, are computed for each subcarrier of the uplink data inside a slot. Matrix-vector multiplication also employs the Intel MKL function ``cblas\_cgemm'' in Figure \ref{fig6}. In this regard, $transa$ should be ``CblasConjTrans'' to realize conjugate transpose multiplication. Intel MKL function ``LAPACKE\_cposv'' is used to solve for X in the linear equation \eqref{eq22}. This routine solves for $X$ in real or complex systems of linear equation ${AX = B}$, where $A$ is an $n$-by-$n$ symmetric/Hermitian positive-defined matrix, the columns of matrix $B$ are individual right-hand sides, and $X$ columns are the corresponding solutions. Cholesky decomposition involves factor $A$ as ${A=LL^{\rm{H}}}$ (complex flavors) when ${uplo = L}$, where $L$ is a low triangular matrix, and the factored form of $A$ is used to solve the system of equation ${AX = B}$ [9]. Finally, recovered symbol are stored in a regulated format, such as that in Figure \ref{fig4}.

\section{Numerical and Scenario Test Results}
This section discusses simulation and scenario test results to validate the performance and feasibility of the uplink receiver in LTE-like TDD-based 16$\times$4 MU massive MIMO systems.

\subsection{Simulation Results}
Monte Carlo method is utilized to simulate on MATLAB based on 14400000 bits generated randomly and to compare the MSE performance among the aforementioned channel estimation schemes. Table \uppercase\expandafter{\romannumeral3} tabulates the system simulation parameters.

Figure \ref{fig7} shows the simulation results. Figure \ref{fig7}(a) presents MSE performances under accurate synchronization. The 12W-LMMSE channel estimator offers approximately 7 dB gain over the LS method and is close to that of the LMMSE estimator with a difference of 3 dB. The 3W-LMMSE channel estimator presents approximately 3.8 dB gain over LS and 2 dB gain over three-fold interpolated LS.

Figure \ref{fig7}(b) displays the MSE performance when a two-point timing synchronization error exists. A two-point timing synchronization error leads to a slight degradation on the MSE performance of MMSE-based estimators. Therefore, the 12W-LMMSE and 3W-LMMSE channel estimators are robust in detecting several point timing errors, and this property is crucial to implementation of prototyping testbeds because a practical system cannot ensure accurate synchronization at all times.

\begin{table*}
	\renewcommand{\arraystretch}{1.3}
	\caption{Comparision of time cost (per core) for estimators: 
		\small \ding{172} LS, \ding{173} 3-fold interpolated LS, \ding{174} 3W-LMMSE, \ding{175} 12W-LMMSE}
	\centering
	\begin{tabular}{p{1.5cm}<{\centering}p{4cm}<{\centering}p{3cm}<{\centering}p{4cm}<{\centering}p{3.5cm}<{\centering}}	
		\toprule[1.2pt]
		\textbf{Methods} & \textbf{Estimation cycles (time)} & \textbf{Total cycles (time)} & \textbf{Duty cycle (theoretical)} & \textbf{Duty cycle (realistic)}\\
		\hline
		\ding{172} & $3.28 \times {10^4}({\rm{0}}.{\rm{01ms}})$ & $2.99 \times {10^6}(1.07{\rm{ms}})$  & ${\rm{10}}{\rm{.7\% }}$ &  ${\rm{12}}{\rm{.1\% }}$ \\
		\ding{173}  & $2.12 \times {10^5}(0.08{\rm{ms}})$ & $3.50 \times {10^6}(1.25{\rm{ms}})$ &  ${\rm{12}}{\rm{.5\% }}$ & ${\rm{13}}{\rm{.1\% }}$ \\
		\ding{174} & $3.55 \times {10^6}(1.27{\rm{ms}})$ & $6.34 \times {10^6}(2.26{\rm{ms}})$ & ${\rm{22}}{\rm{.6\% }}$ & ${\rm{21}}{\rm{.6\% }}$  \\
		\ding{175} & $1.36 \times {10^7}(4.86{\rm{ms}})$ & $2.27 \times {10^7}(8.11{\rm{ms}})$ & ${\rm{81}}{\rm{.1\% }}$ & ${\rm{85}}{\rm{.3\% }}$ \\
		\bottomrule[1.2pt]
		\hline
	\end{tabular}
\end{table*}

MSE performance is re-evaluated when parameters ${L}$ and $ {\rm{SNR_{fixed}}} $ are mismatched with realistic values (still with a two-point timing error). 

Figure \ref{fig7}(c) illustrates the MSE performance of MMSE-based channel estimators along with increasing assumed channel taps $L$ from 1 to 15 when several channel taps exist. LMMSE possesses the least MSE when $L$ equals the real tap number 7. When $L$ is less than 7 and distant from the correct value, MSE is large, and this result is undesirable. These results are attributed to some neglected channel paths and discarded carried information. As a result of additional noise, performance decreases slightly when $L$ is slightly higher than 7. This trend is also suitable for 12W-LMMSE and 3W-LMMSE although the degree of performance decrease is less when less subcarrier correlation is considered. These observations imply that MSE performance of LMMSE-based channel estimation methods can be easily affected when $L$ is less than the real value, especially when more subcarriers are involved. On the other hand, LMMSE-based channel estimation schemes show robustness when $L$ is set slightly higher than the real number of taps. 

Figure \ref{fig7}(d) shows that MSE performance of MMSE-based channel estimators improves with the increase in assumed $ {\rm{SNR_{fixed}}} $ from 0 dB to 40 dB and when realistic $ {\rm{SNR} = 25 dB} $ dB. Thus, MMSE-based channel estimators are robust when the assumptive $ {\rm{SNR_{fixed}}} $ is higher than the realistic $ {\rm{SNR}} $.



\subsection{Scenario Test Results}
Two scenario tests are conducted to evaluate the performance of a multi-core GPP-based uplink receiver in terms of BER performance and time cost.

The \textbf{indoor test scenario} occurs in a typical office room with an area of approximately 30 m$^2$, as shown in Figure \ref{fig8}(a). The \textbf{corridor test scenario} is in a long narrow corridor approximately 25 m long, as shown in Figure \ref{fig8}(b). The component of RaPro can be observed directly in these images. Four single-antenna UEs are connected to the four interfaces of two UE SDRs encapsulated in the box, and wireless signals are sent to 16 BS antennas. UE data are processed by BS SDRs, collected in the PXI chassis, and sent to the server with multi-core GPPs through two SPF+ cables. Figure \ref{fig8}(c) displays the details of some components. The distance between transmitting and receiving antennas approximates 1.5 m, and modulation schemes of the four UEs comprise QPSK, 64-QAM, 16-QAM, and 64QAM.

To refine $L$, BER performance of UE 2 under different $L$ is measured with 64-QAM and MMSE detection. Table \uppercase\expandafter{\romannumeral4} provides the results under indoor and corridor test scenarios. Antenna gain is configured as 30 dB. BER of 12W-LMMSE declines from $L=5$ to $L=8$ in an indoor scenario but rises in the corridor scenario. This difference suggests that channel tap numbers of indoor and corridor scenarios are equal to and less than 8, respectively. This result is reasonable as the number of reflection paths in open places, such as the corridor scenario, is less than that in an office filled with reflectors.

Compared with the indoor scenario, BER in corridor scenario presents a larger gain of 12W-LMMSE over 3W-LMMSE, indicating that 12W-LMMSE channel estimation outperforms 3W-LMMSE in a corridor scenario than in an indoor scenario. BER of 3W-LMMSE and 12W-LMMSE channel estimation presents a rising trend with increasing $L$, and this finding differs from MSE simulation results shown in Figures \ref{fig7}(c) and \ref{fig7}(d). These observations may be due to the difference between the simulation and realistic channels. We adopt $L = 7$ in both scenarios because it is a moderate value, and BER performance is relatively stable between 5 and 10.

Figures \ref{fig9}(a) and \ref{fig9}(b) show BER performances of UE 2 with $L = 7$ 64-QAM and MMSE detection in indoor and corridor scenarios, respectively. 12W-LMMSE channel estimation performs best and is followed by 3W-LMMSE, three-fold interpolated LS, and an LS channel estimator; this result is the same as that in a previously indicated MSE performance simulation. The designed low-complexity channel estimation mapping rules 12W-LMMSE and 3W-LMMSE result in prominent BER performance improvement because their BER can reach approximately 100-fold and 10-fold lower than 3-fold interpolated LS and LS channel estimators when antenna gain measures 30 dB. To reach $10^{-3}$ BER value (without channel coding), a 12W-LMMSE channel estimator can save approximately 4, 9, and 11 dB antenna gains compared with 3W-LMMSE, 3-fold interpolated LS, and LS channel estimators, respectively.

A corridor scenario achieves better BER performance compared with an indoor scenario when antenna gain totals more than 25 dB for 12W-LMMSE channel estimation. A 12W-LMMSE channel estimator achieves $8\times10^{-5}$ under an indoor scenario with 30 dB antenna gain, whereas it can achieve $10^{-6}$ under a corridor scenario.

BER performance of 3W-LMMSE under a corridor scenario is similar to that in an indoor case, whereas 12W-LMMSE under a corridor case performs better than that in indoor case, agreeing with the former analysis that 12W-LMMSE performs better under a corridor scenario than an indoor scenario.

Table \uppercase\expandafter{\romannumeral5} summarizes the time costs. Estimation cycles represent the average clock cycles consumed to conduct channel estimation of each slot processing thread. Estimation time is the corresponding time that considers the principal frequency at 2.8 GHz. The total mean cycle (time) refers to the average cycles (time) for figuring out the data bits of each slot processing thread. Given that 18 slot processing threads run in parallel, the total cycles are equivalent to the time cycles in processing data within a whole 10 ms frame. Estimation cycles of 3W-LMMSE and 12W-LMMSE account for more than half of the total cycles, suggesting the high complexity of MMSE-based channel estimation. The theoretical duty cycle is calculated by dividing the total time by 10 ms, which is close to the realistic duty cycles shown on the system monitor: ${\rm{12}}{\rm{.1\% }}$, ${\rm{13}}{\rm{.1\% }}$, ${\rm{21}}{\rm{.6\% }}$, ${\rm{85}}{\rm{.3\% }}$. 

Table \uppercase\expandafter{\romannumeral5} shows that realistic duty cycles are all less than ${\rm{100}}{\rm{\% }}$. Therefore, time costs of a multi-core GPP-based uplink receiver that contains a 3W-LMMSE or 12W-LMMSE channel estimator and an MMSE MIMO detector allow real-time transmission of RaPro.


\section{Conclusions}
In this paper, we applied the massive MIMO uplink receiver on multi-core GPPs of an MU massive MIMO 5G rapid prototyping system RaPro. We also present the overall multi-threads and functional design and discuss specific procedures of 12W-LMMSE channel estimation and a MMSE MIMO detector. Numerical simulation results indicate the robustness and standard parameter of choice for the designed low-complexity LMMSE channel estimation schemes 3W-LMMSE and 12W-LMMSE. Indoor and corridor scenario test results suggest the feasibility and practicality of this uplink receiver implementation.

\end{document}